\documentclass[]{Liebert_Author_revised}
\usepackage{graphicx}
\usepackage{multicol}
\usepackage{multirow}

\usepackage{url}
\usepackage{color}
\usepackage{marvosym}
\usepackage{float}

\usepackage{caption}
\usepackage{subcaption}
\usepackage{cite}
\usepackage{amsmath,amssymb,amsfonts}
\usepackage{algorithmic}

\newcommand{\eg}{\emph{e.g.}}

\title{
Uncovering Hierarchical Structure in LLM Embeddings Using Geometric and Topological Measures
}

\author{Prakash Chourasia$^1$, Sarwan Ali$^2$, Murray Patterson$^1$, and Alex Zelikovsky$^1$\\
  {$^1$Georgia State University} 
  {Atlanta, GA 30303, USA}\\
  {$^2$Columbia University, New York, NY, USA}\\
E-mail: pchourasia1@student.gsu.edu, sa4559@cumc.columbia.edu, \\
   mpatterson30@gsu.edu, azelikovsky@cs.gsu.edu \\ 
}

\date{}

\begin{document} 

\maketitle 


\begin{abstract}
The rapid advancement of large language models (LLMs) has driven major advances across scientific domains, yet the hierarchical, geometric, and topological structure of their embedding spaces remains poorly understood. In this work, we introduce a comprehensive framework for evaluating LLM embeddings through six complementary metrics that characterize hierarchy, geometry, and topology. Specifically, $\delta$-hyperbolicity and ultrametricity quantify the degree of tree-like structure, while Neighbor Joining evaluates how well pairwise distances support hierarchical reconstruction. The Ollivier-Ricci curvature captures local geometric behavior, persistent homology measures multiscale topological stability, and fractal dimension estimates intrinsic complexity. Applying this framework to embeddings from multiple protein LLMs, we reveal how different models encode hierarchical relationships and structural organization. Evaluating the structure of these embeddings in the representation space helps improve geometry-aware model design. This is critical in domains such as bioinformatics and healthcare, where real-world hierarchical and relational structures matter. Our results demonstrate that geometric and topological analysis provides a guiding lens for understanding and potentially improving the representational quality of LLM embeddings.
\end{abstract}

\section{Introduction}
\label{sec_intro}
The integration of large language models (LLMs) in healthcare has advanced rapidly. These models generate embeddings that capture rich contextual and sequential information, enabling downstream tasks such as protein function prediction and structure inference. This mapping of proteins in hyperbolic space~\citeauthor{dress2012basic} enables efficient clustering of functionally similar proteins, thus supporting applications in Functional Annotation of Proteins~\citeauthor{ali2023protein}. However, understanding the geometric properties of the embeddings is crucial to improving reliability~\citeauthor{nickel2017poincare}.
One way to assess the structural properties of embeddings is through $\delta$-hyperbolicity, a concept introduced by Gromov~\citeauthor{gromov1987hyperbolic}. $\delta$-hyperbolicity quantifies the extent to which a metric space deviates from being an ideal tree~\citeauthor{yang2024enhancing} like structure, where lower $\delta$ values indicate structures closer to a tree-like geometry, which suggests hierarchical relationships among data points. 
In contrast, a higher  $\delta$ indicates a more complex, non-hierarchical structure. 
A major empirical study tested whether social and information networks are tree-like using Gromov’s $\delta$-hyperbolicity~\citeauthor{adcock2013tree}, showing that although traditional metrics miss such structure, hyperbolic analysis reveals meaningful hierarchical patterns.
$\delta$-hyperbolicity quantifies approximate tree-likeness in complex biological networks, simplifying tasks such as distance estimation, network classification, and core–periphery identification~\citeauthor{abu2016metric}. Ultrametricity~\citeauthor{chierchia2019ultrametric} is another popular stricter metric for capturing hierarchical structure, ideal for deterministic, nested data such as phylogenetic trees. In techniques like UPGMA (Unweighted Pair Group Method with Arithmetic Mean) and WPGMA (Weighted Pair Group with Arithmetic Mean), which are used to build ultrametric trees in hierarchical clustering, the idea of ultrametricity is fundamental~\citeauthor{dubes1999cluster}. Whereas Q-matrix statistics from Neighbor Joining offer a quantitative, algorithmic evaluation of tree-likeness that complements these geometric metrics. Strong adherence to additive tree structure is indicated by low normalized Q values~\citeauthor{gascuel2006neighbor,atteson1999performance}. NJ evaluates whether pairwise distances are compatible with an additive tree by transforming the distance matrix into a Q-matrix, demonstrating how strongly each pair of points supports hierarchical merging. 

Beyond tree-likeness, embedding geometry can exhibit additional structural properties that strictly hierarchical metrics do not capture. We incorporate three complementary geometric and topological measures that probe different, orthogonal aspects of the space. These include Ollivier–Ricci curvature~\citeauthor{ollivier2009ricci, ni2015ricci} for local geometric behavior, persistent homology~\citeauthor{edelsbrunner2008persistent} for multiscale topological structure, and fractal dimension~\citeauthor{grassberger1983measuring} for intrinsic geometric complexity. 

We adapt $\delta$-hyperbolicity, ultrametricity, neighborhood joining, and three complementary geometric and topological measures to the domain of genomic sequencing data analyses by evaluating the embeddings generated by LLMs such as ``Tasks Assessing Protein Embeddings (TAPE)"~\citeauthor{rao2019evaluating}, ESM2~\citeauthor{lin2023evolutionary}, Seqvec~\citeauthor{heinzinger2019modeling}, and ProtT5~\citeauthor{elnaggar2021prottrans}. We aim to provide a deeper understanding of the geometric properties underlying LLM-generated representations.
The key contributions of this study are as follows:
\begin{enumerate}
\item We generate embeddings for three datasets using different LLMs and evaluate them with six complementary geometric and topological metrics spanning hierarchy, local geometry, topology, and intrinsic complexity.
\item We provide a comparative geometric analysis of the embeddings of several protein LLMs (SeqVec, ESM2, TAPE, ProtT5), showing how the model design shapes the embedding structure.
\item We assess embedding quality through classification and clustering, demonstrating that geometric properties correlate with downstream performance.
\end{enumerate}


\section{Related Work}
\label{sec_rw}



LLMs have become widely used to produce rich embeddings that boost performance across diverse tasks. Similar models are now used in protein biology to map sequence and structure for better function prediction~\citeauthor{ali2023protein}.
$\delta$-hyperbolicity, introduced by Gromov~\citeauthor{gromov1987hyperbolic}, provides a geometric measure of how tree-like a metric space is. It has been widely applied in graph theory and computational geometry~\citeauthor{krioukov2010hyperbolic} to analyze structural properties relevant to computational problems.
Authors in~\citeauthor{chami2020trees} show that H\textsc{yp}HC closely approximates the optimal tree and outperforms other clustering methods.
Authors in~\citeauthor{chierchia2019ultrametric} frame hierarchical clustering as an optimization problem, seeking the ultrametric that best fits the data. 

Recent work shows that geometric properties of embeddings, including hyperbolicity, can inform clustering and classification~\citeauthor{yang2024enhancing}, but these ideas have not been applied to biological sequence embeddings. Since many biological processes follow tree-like evolutionary patterns~\citeauthor{felsenstein2004inferring, abu2016metric}, $\delta$-hyperbolicity provides a natural tool for assessing and improving models in phylogenetics, viral evolution, and epidemiology. Hyperbolic embedding techniques have also shown benefits in drug–target prediction and functional protein annotation~\citeauthor{dress2012basic, ali2023protein}, highlighting the broader potential of hyperbolic geometry in biological modeling.
Recent studies use geometric tools such as Ricci curvature, persistent homology, and fractal dimension to analyze neural representations~\citeauthor{ni2015ricci, carlsson2009topology, facco2017estimating}. We incorporate these ideas to provide a unified geometric and topological assessment of protein LLM embeddings.

\section{Proposed Approach}
\label{sec_PAa}

In this section, we outline our framework for evaluating embeddings generated by large language models (LLMs). Our analysis proceeds as follows :
\begin{enumerate}
    \item Generate embeddings using the different LLMs.
    \item Compute Euclidean and Poincaré distance matrices.
    \item Evaluate tree-like structure using three complementary metrics: $\delta$-hyperbolicity, ultrametricity, and Neighbor Joining (via the distribution of Q-matrix).
    \item Assess additional geometric and topological structure using three complementary metrics.
    \item Analyze the classification and clustering results to understand the geometric properties of the embeddings.
\end{enumerate}

\subsection{LLMs for embedding generation}
We employ four LLMs, including SeqVec, ESM-2, TAPE, and ProtT5, to generate embeddings and evaluate how well they capture structural patterns in biological sequences.
\paragraph{Seqvec~\citeauthor{heinzinger2019modeling}}
SeqVec is an ELMo-inspired protein language model that learns contextual embeddings directly from large unlabeled protein datasets such as UniRef50~\citeauthor{suzek2015uniref}. It generates fast, alignment-free representations that capture meaningful sequence features and can rival traditional evolutionary methods~\citeauthor{sarzynska2021detecting}.

\paragraph{ESM2~\citeauthor{lin2023evolutionary}}
ESM-2 is a large transformer-based protein language model~\citeauthor{lin2023evolutionary} that predicts structural features directly from single sequences, eliminating the need for MSAs. Scaled to billions of parameters, it captures detailed structural signals~\citeauthor{rao2020transformer} and powers fast, high-accuracy tools like ESMFold, advancing single-sequence structure prediction.

\paragraph{TAPE~\citeauthor{rao2019evaluating}}
TAPE is a protein representation learning framework that trains transformer, LSTM, and ResNet models on large unlabeled datasets like UniRef50 using self-supervised objectives. It produces contextual embeddings for each sequence and provides the TAPE benchmark, which standardizes downstream tasks to consistently evaluate how well these representations capture structural and functional properties of proteins.

\paragraph{ProtT5~\citeauthor{elnaggar2007prottrans}}
ProtT5 adapts the T5 transformer to protein sequences, training with a BERT-style masked language modeling objective and using only the encoder for efficient inference~\citeauthor{elnaggar2007prottrans}. Without requiring evolutionary data or MSAs, its embeddings capture significant biophysical characteristics and perform well on tasks such as secondary structure and localization prediction.

\subsection{Distance Matrix Calculation}

We compute pairwise distance matrices for all embeddings using both Euclidean and Poincaré metrics to obtain $N \times N$ distance matrices, where $N$ is the total number of protein sequences. Let $X$ be the set of embeddings.  

For the Euclidean metric, the distance between two embeddings $x_i$ and $x_j$ is given by:
\begin{equation}
\scriptsize
d_E(x_i, x_j) = \sqrt{\sum_{k=1}^d (x_{i,k} - x_{j,k})^2},
\end{equation}
where $d$ is the embedding dimensionality.  

For the Poincaré metric~\citeauthor{nickel2017poincare}, the hyperbolic distance between two sequence embeddings $x_i$ and $x_j$ within the unit ball is computed as:
\begin{equation}
\scriptsize
d_P(x_i, x_j) = \mathrm{arcosh} \left( 1 + 2 \frac{\|x_i - x_j\|^2}{(1 - \|x_i\|^2)(1 - \|x_j\|^2)} \right).
\end{equation}
Both metrics capture complementary geometric properties. Euclidean measures standard linear distances. Poincaré captures hierarchical, tree-like relationships in hyperbolic space.




\section{Geometric and Topological Measures}
\label{sec_evaluation_metrics}

We evaluate embedding structure using six metrics spanning hierarchy, geometry, topology, and complexity: $\delta$-hyperbolicity, ultrametricity, Neighbor-Joining, Ollivier–Ricci curvature, persistent homology, and fractal dimension.

\subsection{$\delta$-hyperbolicity}
The proposed algorithm for evaluating $\delta$-hyperbolicity is outlined in Algorithm~\ref{alg_hyperbolicity}  and is implemented to assess the geometric structure of LLM-derived protein sequence embeddings.
The algorithm begins by taking as input a set of embeddings \(X\) computed from different LLMs, a distance function \(d\), and a user-defined number of samples to evaluate. The goal is to calculate the $\delta$-hyperbolicity value \(\delta\) for the metric space defined by the embeddings. Initially, the embeddings are flattened and padded to a uniform length, then optionally reduced in dimensionality via PCA to retain 99\% of variance, ensuring efficient computation.
First, it computes the pairwise distance matrix \(D\) using the distance function \(d\) over the embeddings, where each entry in \(D\) represents the distance between a pair of embeddings. 
The core of the algorithm involves sampling a specified number of random quadruples \((a, b, c, w)\) from the embedding space, rather than iterating exhaustively over all combinations, which helps it to scale to large datasets.  
To evaluate $\delta$-hyperbolicity, we need to compute the Gromov products for four points \(a\), \(b\), \(c\), and \(w\). The Gromov product \([a, b]_w\) is defined as (also mentioned in~\citeauthor{yang2024enhancing}):

\begin{equation}
\label{gromov_dist_eq1}
\scriptsize
[a, b]_w = \frac{1}{2} \left(d(a, w) + d(b, w) - d(a, b)\right)
\end{equation}

Where \(d\) is the Euclidean or Poincaré distance between embeddings.

The $\delta$-hyperbolicity is then calculated by checking if the following inequality holds for all sampled quadruples \(a\), \(b\), \(c\), and \(w\)~\citeauthor{yang2024enhancing}:

\begin{equation}
\label{gromov_dist_eq2}
\scriptsize
[a, c]_w \geq \min([a, b]_w, [b, c]_w) - \delta
\end{equation}

We estimate $\delta$-hyperbolicity by computing the minimum value of \(\delta\) that satisfies the above condition for all quadruples in the space.

To estimate the $\delta$-hyperbolicity of LLM-derived embeddings, we sample random quadruples of points in the embedding space and compute how much each quadruple deviates from the ideal tree-like geometry. After flattening and optionally reducing the embeddings with PCA, we compute pairwise distances and evaluate the Gromov products for each quadruple. The slack between these products gives a local estimate of the hyperbolic deviation. Aggregating these values across all samples yields the maximum, mean, and standard deviation of $\delta$, which together quantify how tree-like the embedding space is, where lower values indicate stronger hierarchical structure. The algorithm for computing $\delta$ Hyperbolicity is discussed in detail in the Supplementary Section.

Using parallel processing, these $\delta$ values are calculated for all quadruples sampled, after which the algorithm returns the maximum, average, and standard deviation of $\delta$ in all samples. These values collectively quantify the degree of hyperbolicity in the embedding space. A lower average $\delta$ indicates a more tree-like space, suggesting that the embedding structure may support hierarchical relationships.

\begin{algorithm}[h]
\caption{Compute $\delta$-Hyperbolicity}
\label{alg_hyperbolicity}
\small
\noindent
\begin{tabular}{p{1cm}p{6.8cm}}
\textbf{Input:}  & Set of embeddings \( X \), distance function \( d \), number of sampled quadruples \( S \) \\
\textbf{Output:} & Estimated $\delta$-hyperbolicity values: \( \delta_{\max}, \delta_{\text{avg}}, \delta_{\text{std}} \)
\end{tabular}

\begin{algorithmic}[1]

\STATE Sample \( S \) random quadruples \( (a, b, c, w) \) from \( X \)
\STATE Initialize empty list \( \Delta \leftarrow [\,] \)


\FOR{each quadruple \( (a, b, c, w) \)}
    \STATE Compute Gromov products: \COMMENT{using Eq~\ref{gromov_dist_eq1} and Eq~\ref{gromov_dist_eq2}}
    \[
    [a, b]_w = \frac{1}{2} \left( d(a, w) + d(b, w) - d(a, b) \right)
    \]
    \[
    [b, c]_w = \frac{1}{2} \left( d(b, w) + d(c, w) - d(b, c) \right)
    \]
    \[
    [a, c]_w = \frac{1}{2} \left( d(a, w) + d(c, w) - d(a, c) \right)
    \]

     \STATE Compute temporary slack:
    \[
    \delta_{\text{temp}} = [a, c]_w - \min\left( [a, b]_w, [b, c]_w \right)
    \]


    \STATE Append \( \delta_{\text{temp}} \) to \( \Delta \)
\ENDFOR

\STATE Compute:
\[
\delta_{\max} \leftarrow \max(\Delta), \]
\[
\delta_{\text{avg}} \leftarrow \text{mean}(\Delta), \]
\[
\delta_{\text{std}} \leftarrow \text{std}(\Delta)
\]

\RETURN \( \delta_{\max}, \delta_{\text{avg}}, \delta_{\text{std}} \)

\end{algorithmic}
\end{algorithm}


\subsection{Ultrametricity}

A metric space is said to be \textbf{ultrametric} if the distance function $d$ satisfies the \textit{ultrametric inequality}, also called a stricter version of the triangle inequality~\citeauthor{chierchia2019ultrametric}:

\begin{equation}
    \scriptsize
    d(i, k) \leq \max\{d(i, j), d(j, k)\} \quad \forall i, j, k
\end{equation}

This is a stronger condition than the standard triangle inequality. It enforces a strict hierarchy, making it ideal for modeling perfect tree structures like phylogenetic trees and hierarchical clustering dendrograms.

The algorithm~\ref{alg_ultrametricity_violations} determines whether a given distance matrix \( D \in \mathbb{R}^{n \times n} \) satisfies the ultrametric inequality. Specifically, the ultrametric condition requires that for any triplet of points \( (i, j, k) \), the largest pairwise distance among them must not exceed the maximum of the other two distances. To verify this, the algorithm iterates over all unique sampled combinations of triplets in the distance matrix. For each triplet, it retrieves the corresponding distances \( d_{ij} \), \( d_{ik} \), and \( d_{jk} \), then identifies the largest of the three. It compares this maximum value to the sum of the other two (smaller) distances, adjusted by a small tolerance \( \epsilon \) to account for numerical imprecision. If the maximum distance exceeds this sum by more than \( \epsilon \), the matrix is deemed to violate the ultrametric property, and the algorithm returns \textbf{False} immediately. If all triplets conform to the inequality, it concludes by returning \textbf{True}, confirming that the entire matrix exhibits ultrametricity. This algorithm is particularly valuable in fields such as phylogenetics and hierarchical clustering, where validating the ultrametric nature of data helps ensure the reliability of tree-based representations.

For each triplet, we check whether the largest of the three distances exceeds the other two by more than a small numerical tolerance. A violation is decided on the basis of it. We quantify the degree to which the distance matrix resembles a completely hierarchical ultrametric space by adding the maximum, mean, and standard deviation of these violations across all samples. As anticipated in applications such as phylogenetics and hierarchical clustering, this offers a straightforward method to assess whether the embeddings represent nested or tree-like relationships.

\begin{algorithm}[h]
\caption{Ultrametricity of a Distance Matrix}
\label{alg_ultrametricity_violations}
\noindent
\small
\begin{tabular}{p{1cm}p{7cm}}
\textbf{Input:}  & Distance matrix \(D \in \mathbb{R}^{n \times n}\), sample size \(S\), tolerance \(\epsilon\) \\
\textbf{Output:} & Dictionary containing ultrametricity violation metrics
\end{tabular}

\begin{algorithmic}[1]{}

\STATE Initialize counters: \(\text{num\_violations} \leftarrow 0\), \(\text{total\_triples} \leftarrow 0\), and empty list \(\text{violations} \leftarrow []\)

\STATE Randomly sample \(S\) unique triples \(\{(i, j, k)\}\) from \(\{1, \ldots, n\}\) without replacement

\FOR{each sampled triple \((i, j, k)\)}
    \STATE Extract distances \(d_{ij} = D[i,j]\), \(d_{jk} = D[j,k]\), and \(d_{ik} = D[i,k]\)
    \STATE Compute violation terms:
    \[
    v_1 = d_{ik} - \max(d_{ij}, d_{jk})
    \]
    \[
    v_2 = d_{ij} - \max(d_{ik}, d_{jk})
    \]
    \[
    v_3 = d_{jk} - \max(d_{ij}, d_{ik})
    \]
    \STATE Compute violation \(\nu = \max(v_1, v_2, v_3, 0)\)
    \IF{\(\nu > \epsilon\)}
        \STATE Append \(\nu\) to \(\text{violations}\)
        \STATE Increment \(\text{num\_violations} \leftarrow \text{num\_violations} + 1\)
    \ENDIF
    \STATE Increment \(\text{total\_triples} \leftarrow \text{total\_triples} + 1\)
\ENDFOR

\text{// Compute statistics over violations:}

\STATE $\text{max\_violation} = \max(\text{violations}) \text{ if } \text{violations} \neq \varnothing \text{ else } 0$

\STATE $\text{avg\_violation} = \text{mean}(\text{violations}) \text{ if } \text{violations} \neq \varnothing \text{ else } 0$

\STATE $\text{std\_violation} = \text{std}(\text{violations}) \text{ if } \text{violations} \neq \varnothing \text{ else } 0$

\RETURN \text{max\_violation}, \text{avg\_violation}, \text{std\_violation}

\end{algorithmic}
\end{algorithm}


\subsection{Neighbor Joining}

We use another metric called Neighbor Joining (NJ) to gauge how closely the pairwise distance structure of the embeddings follows a tree-like geometry. We employed a Neighbor Joining (NJ)-based analysis produced from the NJ Q-matrix~\citeauthor{trees1987neighbor}. 

Given an $n \times n$ distance matrix $D$, the NJ algorithm defines the Q-matrix as
\begin{equation}
\scriptsize
Q(i,j) = (n-2)D(i,j) - \sum_{k=1}^{n} D(i,k) - \sum_{k=1}^{n} D(j,k)
\end{equation}
where pairs with the most negative $Q(i,j)$ values are preferentially joined during tree construction. 
Rather than explicitly reconstructing a phylogenetic tree, 
we use the NJ Q-matrix, which transforms the distance matrix into a form that highlights how strongly each pair of points supports hierarchical merging. Pairs with highly negative Q-values are those most consistent with additive tree geometry. This makes the Q-matrix a useful diagnostic tool for assessing tree-likeness directly from distances.

Algorithm~\ref{alg_nj_scores} shows the computation that uses the Neighbor-Joining (NJ) Q-matrix to compute tree-likeness statistics from a given distance matrix. By modifying pairwise distances based on their global relationships, it first creates the Q-matrix. After that, all distinct pairwise Q-values are extracted, converted to absolute values, and their distribution is summarized using the maximum, mean, and standard deviation. The degree to which the data exhibit a hierarchical or tree-like structure is concisely captured by these statistics.

To obtain a global measure of hierarchical consistency, we summarize the distribution of all off-diagonal Q-values by reporting their maximum, mean, and standard deviation after normalization. The lower mean and maximum values indicate that the distance matrix more closely follows the NJ criterion and better approximates a tree metric. Combined with other geometric indicators such as hyperbolicity and curvature, these scores provide a robust measure of how strongly an embedding reflects hierarchical structure. Rather than constructing the full tree, we summarize the distribution of Q-matrix values to capture the embedding's overall consistency with a tree-structured geometry.

\begin{algorithm}[h]
\caption{Neighbor Joining–Based Tree-Likeness Scoring}
\label{alg_nj_scores}
\noindent
\small
\begin{tabular}{p{1cm}p{7cm}}
\textbf{Input:}  & Distance matrix \(D \in \mathbb{R}^{n \times n}\) \\
\textbf{Output:} & Dictionary containing NJ score statistics
\end{tabular}

\begin{algorithmic}[1]{}

\STATE Let \(n \leftarrow\) number of samples in \(D\)

\IF{\(n < 3\)}
    \RETURN \(\text{nj\_max} = 0,\ \text{nj\_avg} = 0,\ \text{nj\_std} = 0\)
\ENDIF

\STATE Compute the Neighbor Joining Q-matrix \(Q \in \mathbb{R}^{n \times n}\):
\[
Q(i,j) = (n-2)D(i,j) - \sum_{k=1}^{n} D(i,k) - \sum_{k=1}^{n} D(j,k)
\]

\STATE Initialize empty list \(\mathcal{Q} \leftarrow [ \ ]\)

\FOR{\(i = 1\) to \(n\)}
    \FOR{\(j = i+1\) to \(n\)}
        \STATE Append \(Q(i,j)\) to \(\mathcal{Q}\)
    \ENDFOR
\ENDFOR

\STATE Normalize Q-values: \(\mathcal{Q} \leftarrow \{\, |q| \mid q \in \mathcal{Q} \,\}\)

\STATE \(\text{nj\_max} \leftarrow \max(\mathcal{Q})\)

\STATE \(\text{nj\_avg} \leftarrow \text{mean}(\mathcal{Q})\)

\STATE \(\text{nj\_std} \leftarrow \text{std}(\mathcal{Q})\)

\RETURN \(\text{nj\_max}, \text{nj\_avg}, \text{nj\_std}\)

\end{algorithmic}
\end{algorithm}

\subsection{Ollivier-Ricci Curvature}
To capture local geometric properties complementary to global hyperbolicity, we computed the discrete Ollivier-Ricci curvature~\citeauthor{ollivier2009ricci} on the k-nearest neighbor graph constructed from embeddings (see Algorithm~\ref{alg_Ricci}). 
For adjacent points $x_i$ and $x_j$, the Ollivier-Ricci curvature is defined as:
\begin{equation}
\scriptsize
\kappa(x_i, x_j) = 1 - \frac{W_1(\mu_i, \mu_j)}{d(x_i, x_j)}
\end{equation}
where $W_1(\mu_i, \mu_j)$ is the Wasserstein-1 distance between probability distributions $\mu_i$ and $\mu_j$ centered at $x_i$ and $x_j$ respectively, typically defined as uniform distributions over their k-nearest neighbors~\citeauthor{ni2015ricci}. Negative curvature ($\kappa < 0$) indicates the local exponential divergence characteristic of hyperbolic geometry and tree-like structure~\citeauthor{sandhu2015graph}, while positive curvature suggests clustering or community structure. We report the mean, standard deviation, minimum, and proportion of edges with negative curvature in the
k-NN graph.


\begin{algorithm}[h!]
\caption{Compute Ollivier-Ricci Curvature}
\label{alg_Ricci}
\small
\noindent
\begin{tabular}{p{1cm}p{7cm}}
\textbf{Input:}  & Embedding matrix $X \in \mathbb{R}^{n \times d}$, number of neighbors $k$ \\
\textbf{Output:} & Mean Ricci curvature $\bar{\kappa}$, standard deviation $\sigma_\kappa$, 
        minimum $\kappa_{min}$, negative ratio $r_{neg}$
\end{tabular}
\begin{algorithmic}[1]
\STATE Construct k-NN graph $G = (V, E)$ from $X$
\STATE Initialize $\kappa\_values \leftarrow []$
\FOR{each edge $(i,j) \in E$}
    \STATE $N_i \leftarrow$ k-nearest neighbors of $x_i$
    \STATE $N_j \leftarrow$ k-nearest neighbors of $x_j$
    \STATE Define uniform distributions: $\mu_i = \text{Uniform}(N_i)$, 
           $\mu_j = \text{Uniform}(N_j)$
    \STATE Compute Wasserstein-1 distance: $W_1 = W_1(\mu_i, \mu_j)$
    \STATE $\kappa(i,j) \leftarrow 1 - \frac{W_1}{d(x_i, x_j)}$
    \STATE Append $\kappa(i,j)$ to $\kappa\_values$
\ENDFOR
\STATE $\bar{\kappa} \leftarrow \text{mean}(\kappa\_values)$
\STATE $\sigma_\kappa \leftarrow \text{std}(\kappa\_values)$
\STATE $\kappa_{min} \leftarrow \min(\kappa\_values)$
\STATE $r_{neg} \leftarrow \frac{\#\{\kappa < 0\}}{|\kappa\_values|}$
\RETURN $\bar{\kappa}, \sigma_\kappa, \kappa_{min}, r_{neg}$
\end{algorithmic}
\end{algorithm}

\subsection{Persistent homology}

We employ 0-dimensional persistent homology~\citeauthor{edelsbrunner2008persistent, carlsson2009topology} to quantify the topological structure of embeddings across scales (see Algorithm~\ref{alg_Persistent}). Persistent homology tracks connected components as we increase the distance threshold $\epsilon$ from 0 to $\infty$. Each connected component has a "birth" time (when it first appears) and ``death" time (when it merges with another component). The persistence lifetime $\ell = \text{death} - \text{birth}$ measures the stability of that topological feature~\citeauthor{chazal2016structure}.
\\
For tree-like structures with clear hierarchical organization, components merge rapidly, yielding short persistence lifetimes. In contrast, flat or diffuse geometries exhibit long persistence lifetimes as components remain separated across many scales. We compute the persistence diagram using the Ripser algorithm~\citeauthor{bauer2021ripser} and report:
\begin{itemize}
\item Mean persistence lifetime: $\bar{\ell} = \frac{1}{n-1}\sum_{i=1}^{n-1} \ell_i$
\item Maximum persistence lifetime: $\max_i \ell_i$
\item Persistence entropy: $H = -\sum_{i=1}^{n-1} p_i \log p_i$ where 
      $p_i = \ell_i / \sum_j \ell_j$
\end{itemize}
Lower persistence values indicate stronger hierarchical consolidation.

\begin{algorithm}[h!]
\caption{Compute Persistent Homology Features}
\label{alg_Persistent}
\small
\noindent
\begin{tabular}{p{1cm}p{7cm}}
\textbf{Input:}  & Distance matrix $D \in \mathbb{R}^{n \times n}$ \\
\textbf{Output:} & Mean persistence $\bar{\ell}$, max persistence $\ell_{max}$, persistence entropy $H$
\end{tabular}
\begin{algorithmic}[1]
\STATE Compute 0-dimensional persistence diagram using Ripser: 
       $\{(\text{birth}_i, \text{death}_i)\}_{i=1}^{n-1}$
\STATE Compute lifetimes: $\ell_i = \text{death}_i - \text{birth}_i$
\STATE $\bar{\ell} \leftarrow \frac{1}{n-1}\sum_{i=1}^{n-1} \ell_i$
\STATE $\ell_{max} \leftarrow \max_{i} \ell_i$
\STATE Normalize: $p_i \leftarrow \frac{\ell_i}{\sum_{j=1}^{n-1} \ell_j}$
\STATE $H \leftarrow -\sum_{i=1}^{n-1} p_i \log p_i$
\RETURN $\bar{\ell}, \ell_{max}, H$
\end{algorithmic}
\end{algorithm}

\subsection{Fractal Dimension}

We estimate intrinsic dimensionality using the correlation dimension~\citeauthor{grassberger1983measuring}, derived from the scaling behavior of the correlation integral for points ${x_i}$. Details are mentioned in Algorithm~\ref{alg_Fractal}.  Correlation integral for points ${x_i}$ can be defined by:
\begin{equation}
\scriptsize
C(r) = \lim_{n \to \infty} \frac{1}{n^2} \sum_{i \neq j} \mathbb{I}(d(x_i, x_j) < r)
\end{equation}
In the scaling region, $C(r) \propto r^{d_f}$, so $d_f$ can be estimated from the slope of $\log C(r)$ vs $\log r$~\citeauthor{camastra2016intrinsic}. We use the two-nearest-neighbor method~\citeauthor{facco2017estimating} for robust estimation:
\begin{equation}
\scriptsize
d_f = \left[ \frac{1}{n} \sum_{i=1}^n \log\left(\frac{r_2(x_i)}{r_1(x_i)}\right) \right]^{-1}
\end{equation}
where $r_1(x_i)$ and $r_2(x_i)$ are the distances to the first and second nearest neighbors of $x_i$. Lower fractal dimension signals embeddings concentration near low-dimensional manifolds, whereas higher values indicate a more diffuse, isotropic spread in high-dimensional space~\citeauthor{levina2004maximum}.

\begin{algorithm}[h!]
\caption{Estimate Fractal Dimension}
\label{alg_Fractal}
\small
\noindent
\begin{tabular}{p{1cm}p{7cm}}
\textbf{Input:}  & Embedding matrix $X \in \mathbb{R}^{n \times d}$ \\
\textbf{Output:} & Mean fractal dimension $\bar{d}_f$, standard deviation $\sigma_{d_f}$, median $\tilde{d}_f$
\end{tabular}
\begin{algorithmic}[1]
\STATE $d_f\_values \leftarrow []$
\FOR{each point $x_i \in X$}
    \STATE Find distances to 2 nearest neighbors: $r_1(x_i), r_2(x_i)$
    \STATE $d_{f,i} \leftarrow \left[\log\left(\frac{r_2(x_i)}{r_1(x_i)}\right)\right]^{-1}$
    \STATE Append $d_{f,i}$ to $d_f\_values$
\ENDFOR
\STATE $\bar{d}_f \leftarrow \text{mean}(d_f\_values)$
\STATE $\sigma_{d_f} \leftarrow \text{std}(d_f\_values)$
\STATE $\tilde{d}_f \leftarrow \text{median}(d_f\_values)$
\RETURN $\bar{d}_f, \sigma_{d_f}, \tilde{d}_f$
\end{algorithmic}
\end{algorithm}

\section{Experimental Setup}
\label{sec_es}
We performed the experiments on a computer running 64-bit Windows 10 with an Intel(R) Core i5 processor at 2.10 GHz and 32 GB of RAM. For experiments, we randomly split the data into 60-10-30\% for training, validation, and testing. We repeat the experiments $5$ times and report the average and standard deviation.
For the sake of reproducibility, we have made code available online~\footnote{\url{https://github.com/pchourasia1/Hyperbolicity_LLM_Embeddings}}.

\subsection{Data Statistics}
\label{sub_sec_data_stats}
We use $3$ real-world protein sequence datasets. The summary of the datasets used for the experimentation is given below:

\subsubsection{PDB186}
We use a benchmark dataset from the literature (PDB186) for predicting DNA-protein binding, a binary classification task~\citeauthor{lou2014sequence}. More formally, the dataset contains DNA-binding and non-binding protein sequences. The research task is to predict whether the DNA and protein sequences will bind. The minimum, maximum, and average sequence length in the data are 64, 1323, and 264.693, respectively. The total number of sequences is 186, of which 93 bind and 93 do not.

\subsubsection{Breast Cancer}
Our Membranolytic anticancer peptides (ACPs) dataset~\citeauthor{Grisoni2019} contains 949 protein sequences annotated for anticancer activity against breast cancer cell lines. The target labels are categorized into ``very active'', ``moderately active'', ``experimental inactive'', and ``virtual inactive'' groups. Table~\ref{tab_data_Stats_BC} shows the distribution in detail.

\begin{table}[!t]
      \centering
      \resizebox{0.7\textwidth}{!}{
         \begin{tabular}{lcccc}
    \toprule
    ACPs Category & Count & Min. & Max. & Average \\
    \midrule \midrule
        Inactive-Virtual & 750 & 8 & 30 & 16.64 \\
        Moderate Active & 98 & 10 & 38 & 18.44 \\
        Inactive-Experimental & 83 & 5 & 38 & 15.02 \\
        Very Active & 18 & 13 & 28 & 19.33 \\
        \midrule \midrule
        Total  & 949 & - & - & -\\
        \bottomrule
    \end{tabular}
    }
    \caption{Dataset statistics for the Breast cancer data. Columns represent the minimum, maximum, and average sequence lengths.}
    \label{tab_data_Stats_BC}
\end{table}

\subsubsection{Lung Cancer}
The Membranolytic anticancer peptides (ACPs) dataset~\citeauthor{Grisoni2019} contains 901 sequences annotated for anticancer activity against Lung cancer cell lines.  The target labels are classified into four groups: ``very active," ``moderately active," ``experimental inactive," and ``virtual inactive" groups. Table~\ref{tab_data_Stats_Lung_C} shows the distribution in detail.

\begin{table}[!t]
      \centering
      \resizebox{0.7\textwidth}{!}{
         \begin{tabular}{lcccc}
    \toprule
    ACPs Category & Count & Min. & Max. & Average \\
    \midrule \midrule
        Inactive-Virtual & 750 & 8 & 30 & 16.64 \\
        Moderate Active & 75 & 11 & 38 & 17.76 \\
        Inactive-Experimental & 52 & 5 & 38 & 14.5 \\
        Very Active & 24 & 13 & 28 & 20.70 \\
        \midrule \midrule
        Total  & 901 & - & - & -\\
        \bottomrule
    \end{tabular}
    }
     \caption{Dataset statistics for the Lung cancer data. Columns represent the min., max., and average lengths of the sequence.}
     \label{tab_data_Stats_Lung_C}
\end{table}

\subsection{Classification}
For the classification task, we use classifiers like SVM, Naive Bayes, MLP, KNN, Random Forest, Logistic Regression, and Decision Tree. Evaluation metrics include accuracy, precision, recall, weighted F1, macro F1, ROC-AUC, and training runtime. 

\subsection{Clustering}
For the clustering task, we use $k$-means~\citeauthor{kmeans}, $k$-modes~\citeauthor{huang-1998-extensions}, and the agglomerative clustering algorithm~\citeauthor{ward1963hierarchical}. The evaluation metrics used were the Silhouette Coefficient~\citeauthor{rousseeuw1987silhouettes}, Calinski-Harabasz Score~\citeauthor{calinski1974dendrite}, and the Davies-Bouldin Score~\citeauthor{davies1979cluster}.

\subsubsection{Algorithms}

\paragraph{$k$-means~\citeauthor{kmeans}}
The $k$-means~\citeauthor{kmeans} clusters
objects based on the Euclidean mean. The $k$-means clustering partitions the data into $K$ groups by repeatedly assigning points to the nearest centroid and updating the centroids as the mean of their assigned points. This continues until the assignments stabilize, resulting in compact, well-separated clusters~\citeauthor{kanungo2002efficient}.

\paragraph{$k$-modes~\citeauthor{huang-1998-extensions}}
We also perform clustering with $k$-modes~\citeauthor{huang-1998-extensions}, which is a variant of $k$-means
using \textit{modes} instead of \textit{means}.  Here, pairs of
objects are subject to a dissimilarity measure (\eg, Hamming distance) rather than to the Euclidean mean.

\paragraph{Agglomerative~\citeauthor{ward1963hierarchical}}
All agglomerative hierarchical clustering methods start from a proximity matrix of either similarities or distances between elements ~\citeauthor{nei1978estimation}. Agglomerative clustering is a bottom-up algorithm that starts with each data point as its own cluster. It repeatedly merges the closest pairs of clusters based on a chosen distance metric until all points are grouped into a single cluster or a set number of clusters~\citeauthor{mojena1977hierarchical}.

\subsubsection{Evaluation Metric}

\paragraph{Silhouette Coefficient~\citeauthor{rousseeuw1987silhouettes}}
Measures how similar a point is to its own cluster versus similarity with other clusters. Scores range from $-1$ to $1$, with higher values indicating better clustering.

\paragraph{Calinski–Harabasz Score~\citeauthor{calinski1974dendrite}}
Evaluates clustering quality using between-cluster and within-cluster dispersion. Higher values indicate more distinct, well-separated clusters.

\paragraph{Davies–Bouldin Score~\citeauthor{davies1979cluster}}
Computes the ratio of within-cluster to between-cluster distances. Lower values indicate better separation between clusters.

\subsection{Visualization}
\begin{figure*}[h!]
     \centering
    \begin{subfigure}[b]{0.23\textwidth}
         \centering
         \includegraphics[scale=0.075]{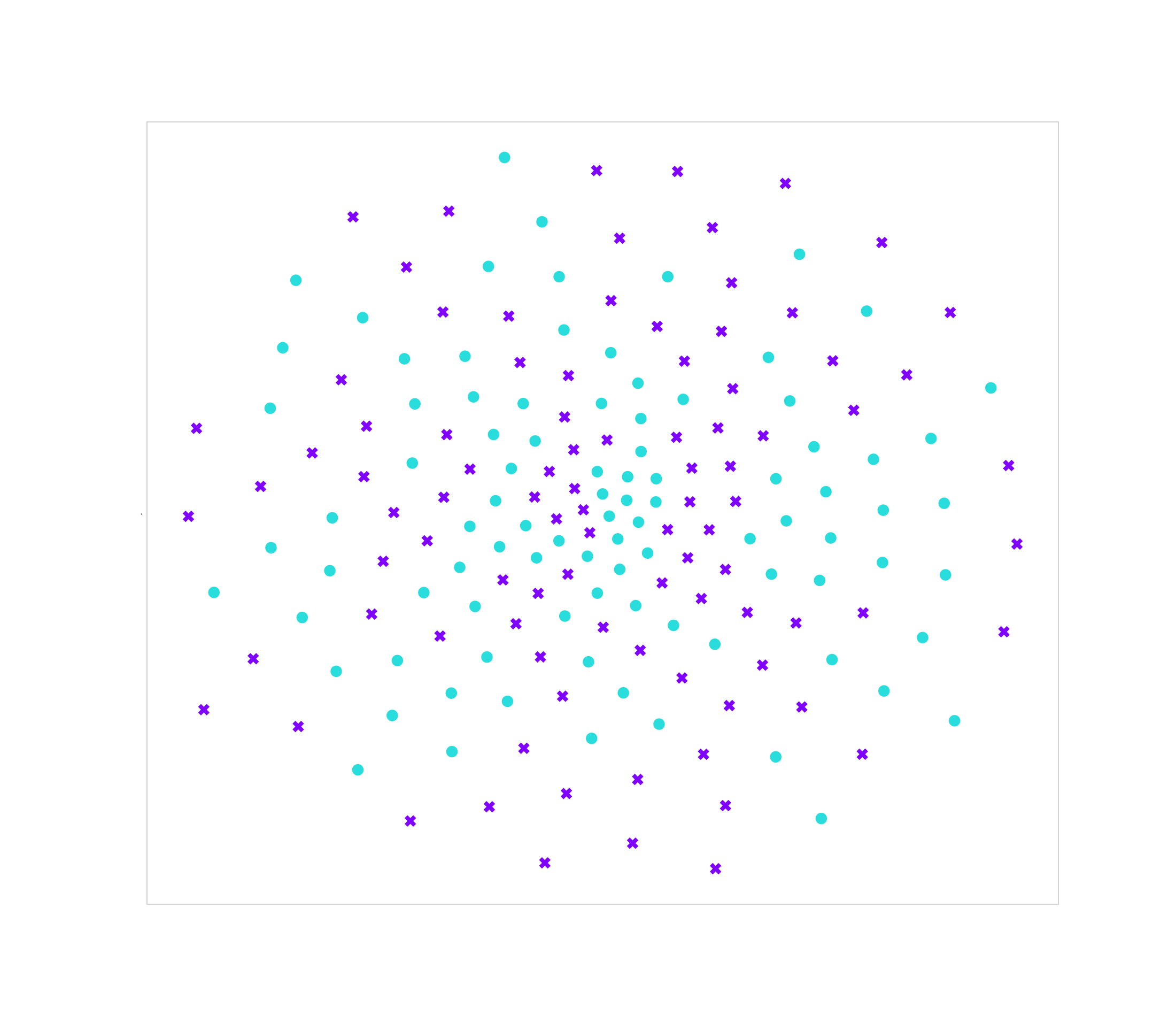}
         \caption{Seqvec}
         \label{tsne_seqvec_pdb186}
     \end{subfigure}
     \hfill
     \begin{subfigure}[b]{0.23\textwidth}
         \centering
         \includegraphics[scale=0.075]{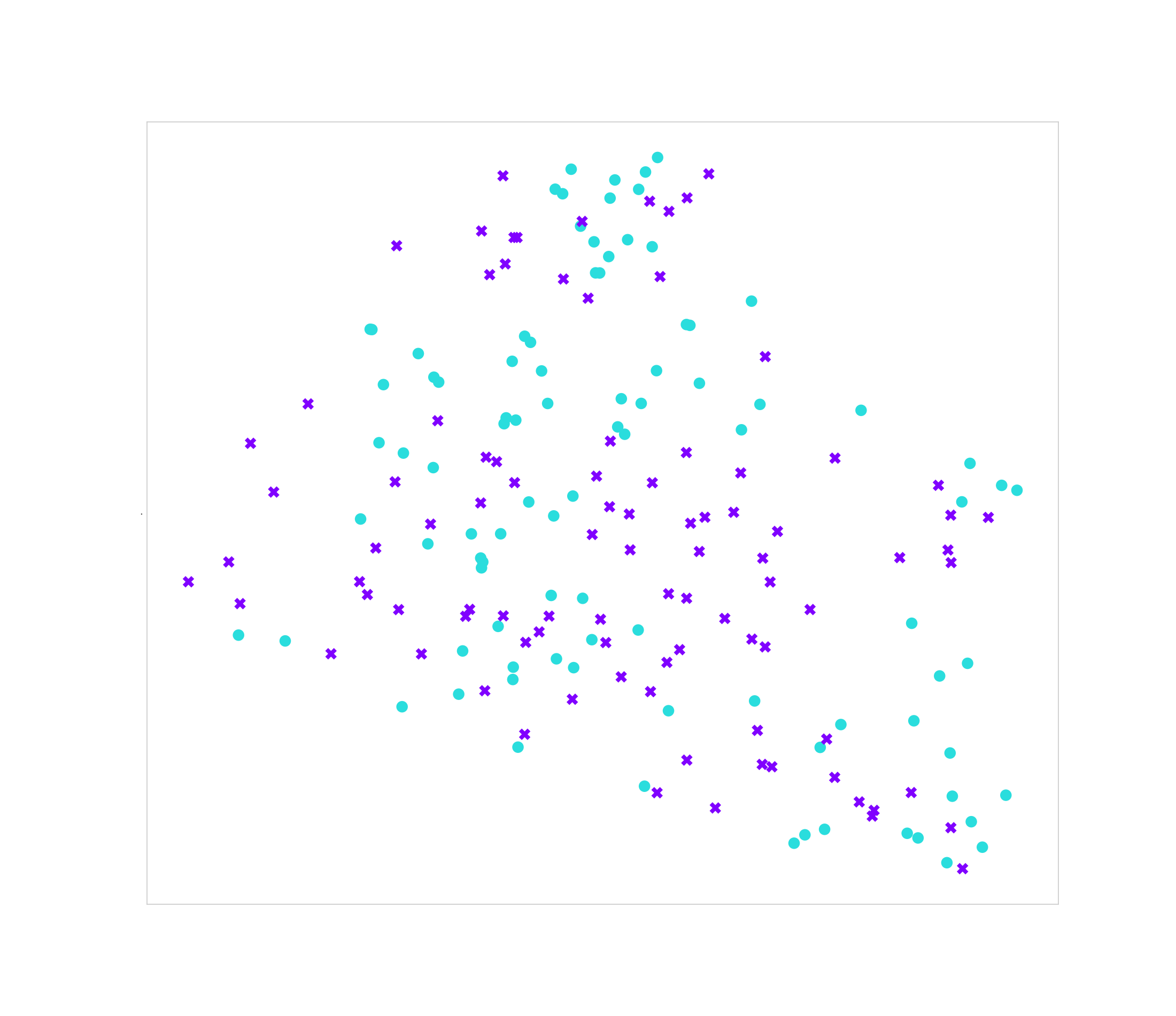}
         \caption{ESM2}
         \label{tsne_esm2_pdb186}
     \end{subfigure}
     \begin{subfigure}[b]{0.23\textwidth}
         \centering
         \includegraphics[scale=0.075]{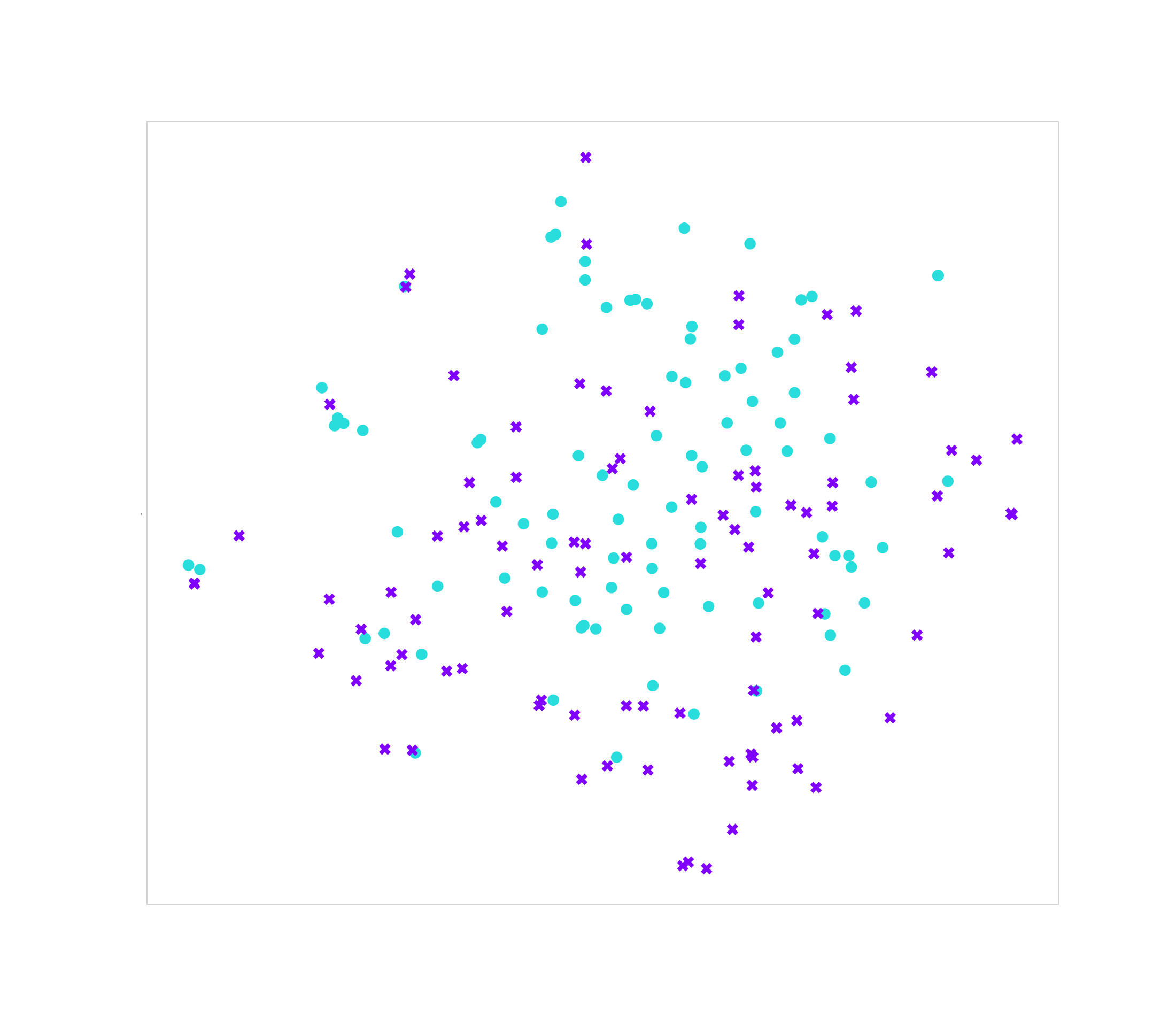}
         \caption{TAPE}
         \label{tsne_tape_pdb186}
     \end{subfigure}
          \hfill
     \begin{subfigure}[b]{0.23\textwidth}
         \centering
         \includegraphics[scale=0.075]{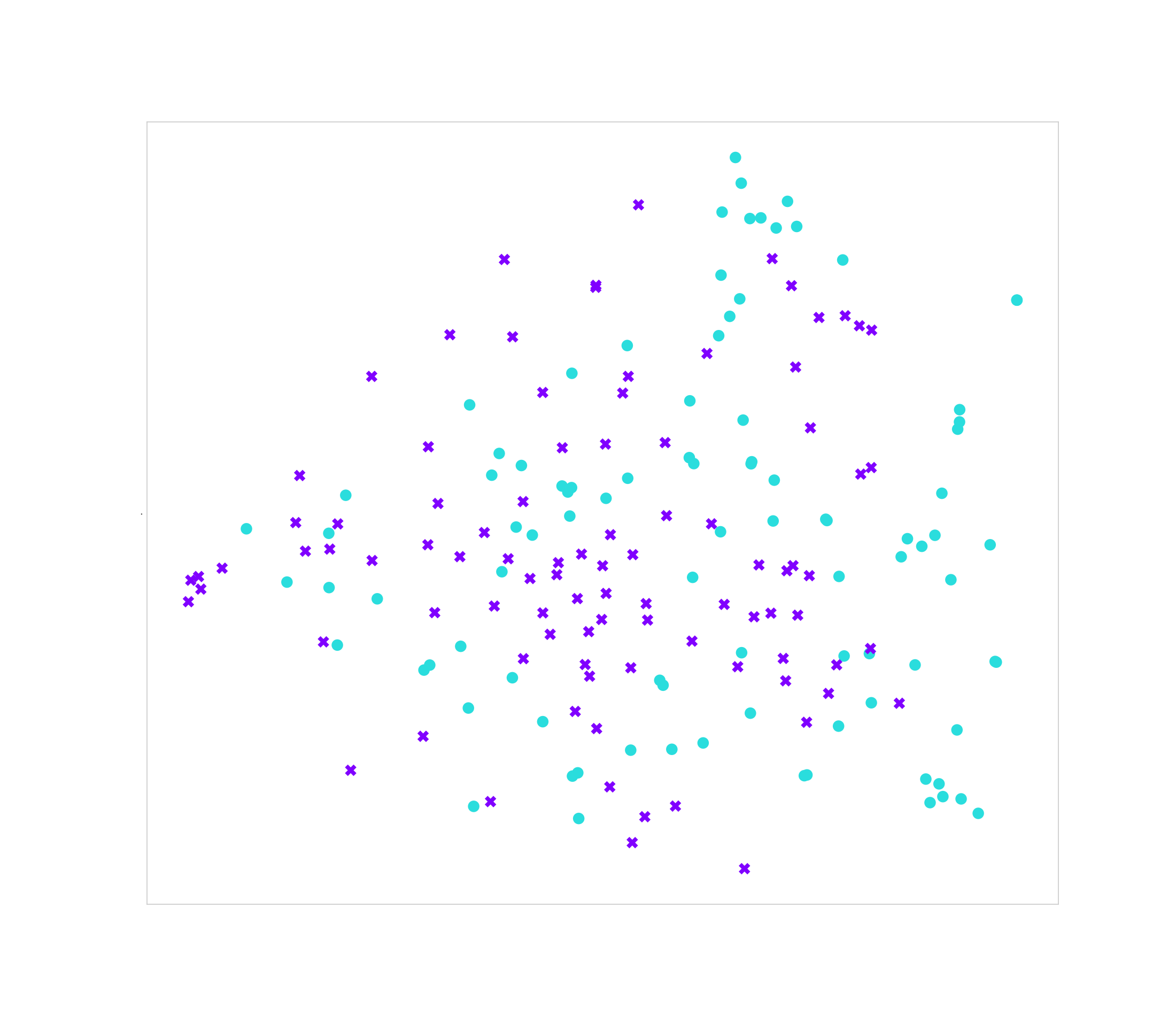}
         \caption{ProtT5}
         \label{tsne_protT5_pdb186}
     \end{subfigure}
     \includegraphics[width=0.65\textwidth]{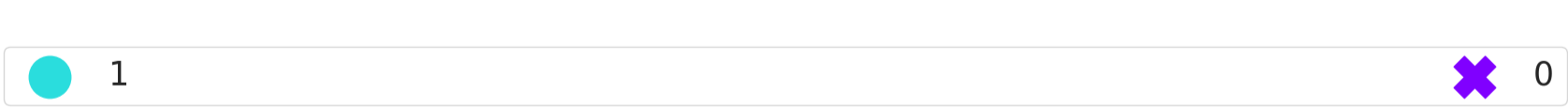}
     \caption{t-SNE visualization of different embeddings from\textbf{ PDB 186 dataset}. The Figure is best seen in color.}
     \label{tsne_pdb186}
\end{figure*}

\begin{figure*}[h!]
     \centering
    \begin{subfigure}[b]{0.23\textwidth}
         \centering
         \includegraphics[scale=0.075]{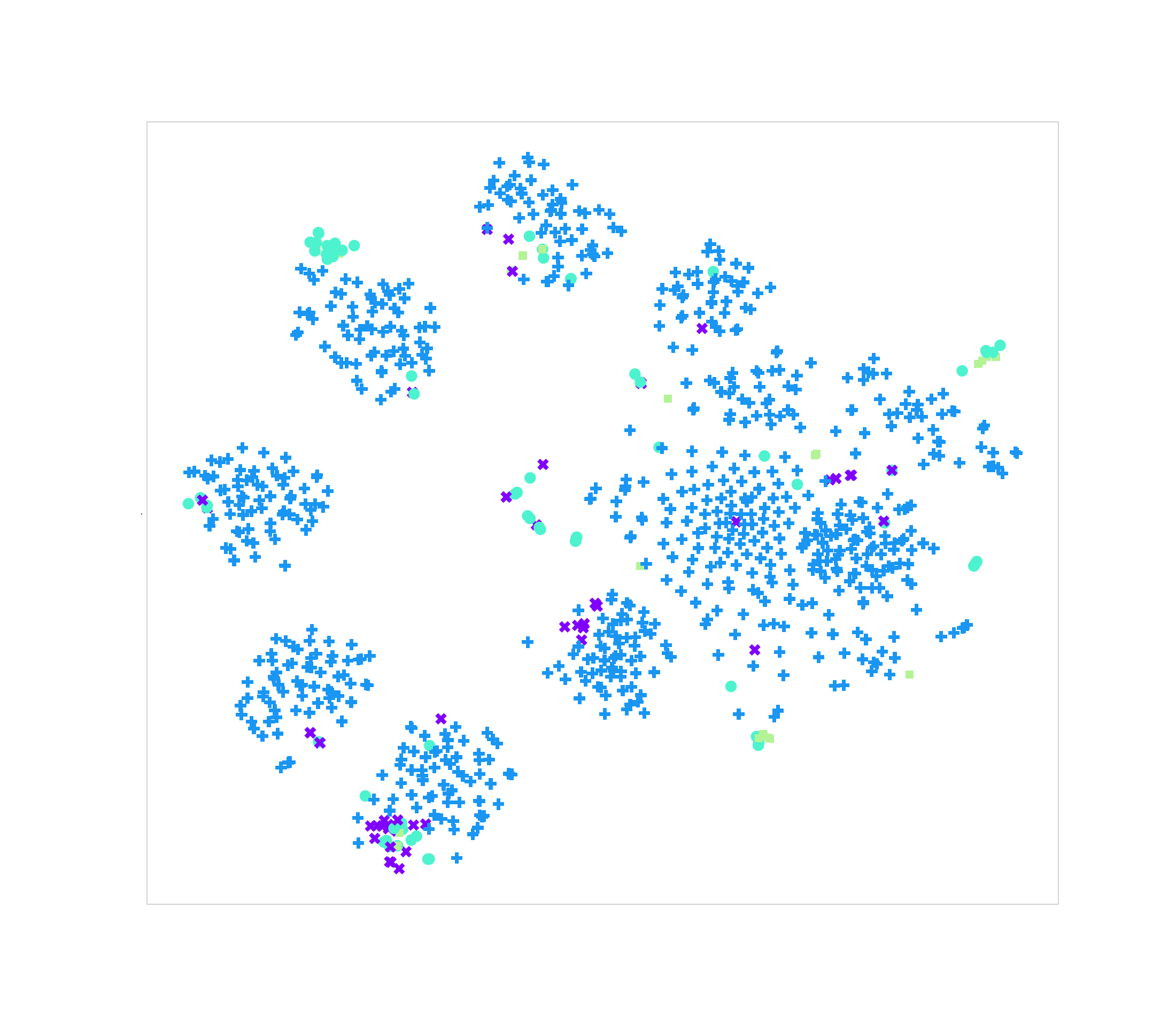}
         \caption{Seqvec}
         \label{tsne_seqvec_LC}
     \end{subfigure}
     \hfill
     \begin{subfigure}[b]{0.23\textwidth}
         \centering
         \includegraphics[scale=0.075]{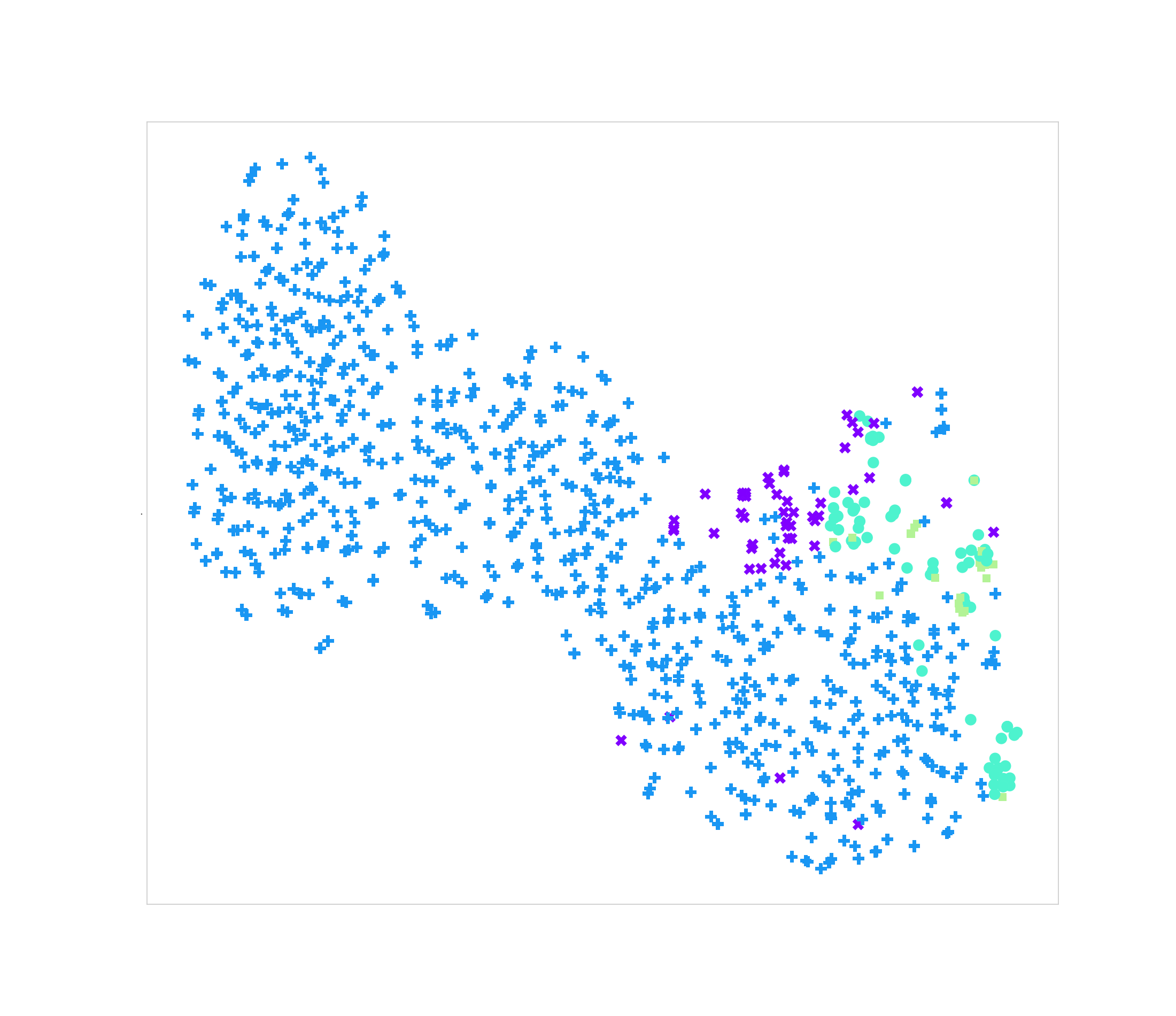}
         \caption{ESM2}
         \label{tsne_esm2_LC}
     \end{subfigure}
     \begin{subfigure}[b]{0.23\textwidth}
         \centering
         \includegraphics[scale=0.075]{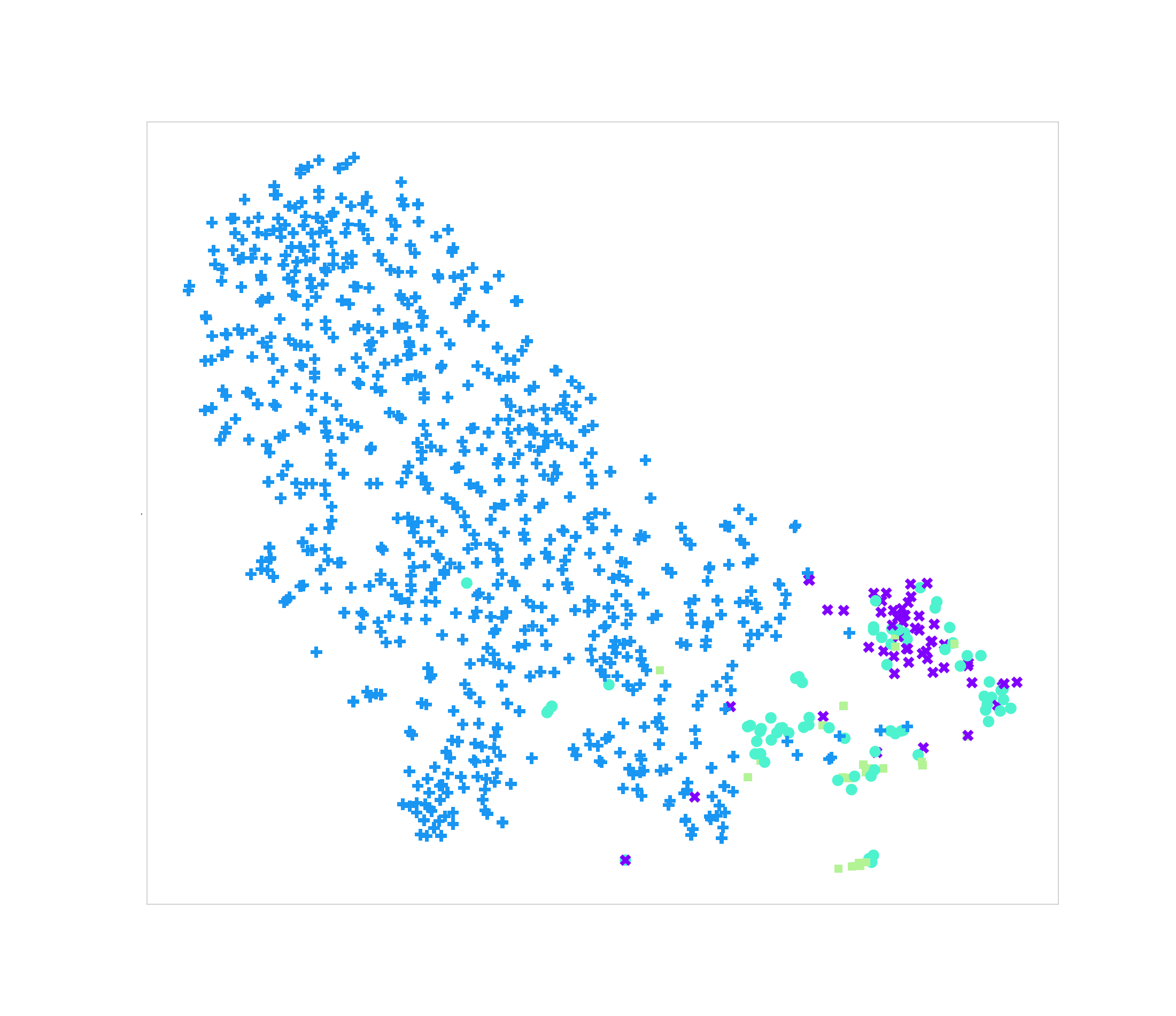}
         \caption{TAPE}
         \label{tsne_tape_LC}
     \end{subfigure}
          \hfill
     \begin{subfigure}[b]{0.23\textwidth}
         \centering
         \includegraphics[scale=0.075]{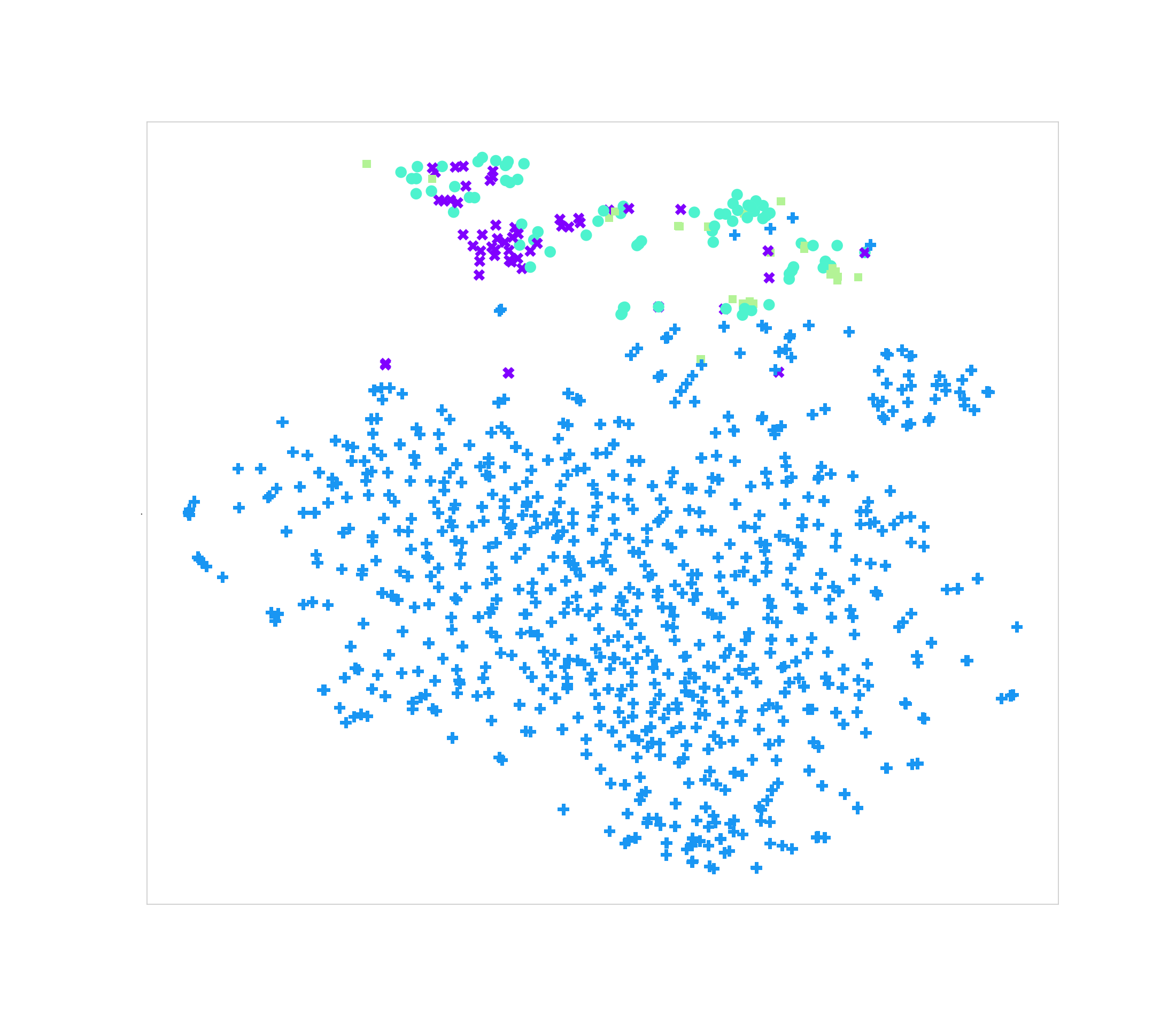}
         \caption{ProtT5}
         \label{tsne_protT5_LC}
     \end{subfigure}
     \includegraphics[width=0.75\textwidth]{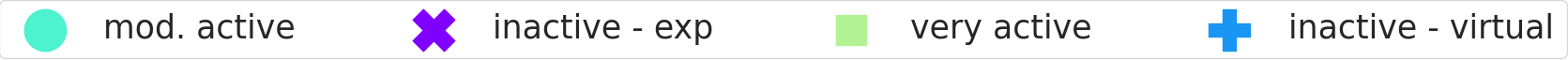}
     \caption{t-SNE visualization of different embeddings from\textbf{ Lung Cancer dataset}. The Figure is best seen in color.}
     \label{tsne_lung_cancer}
\end{figure*}

\begin{figure*}[h!]
     \centering
    \begin{subfigure}[b]{0.23\textwidth}
         \centering
         \includegraphics[scale=0.075]{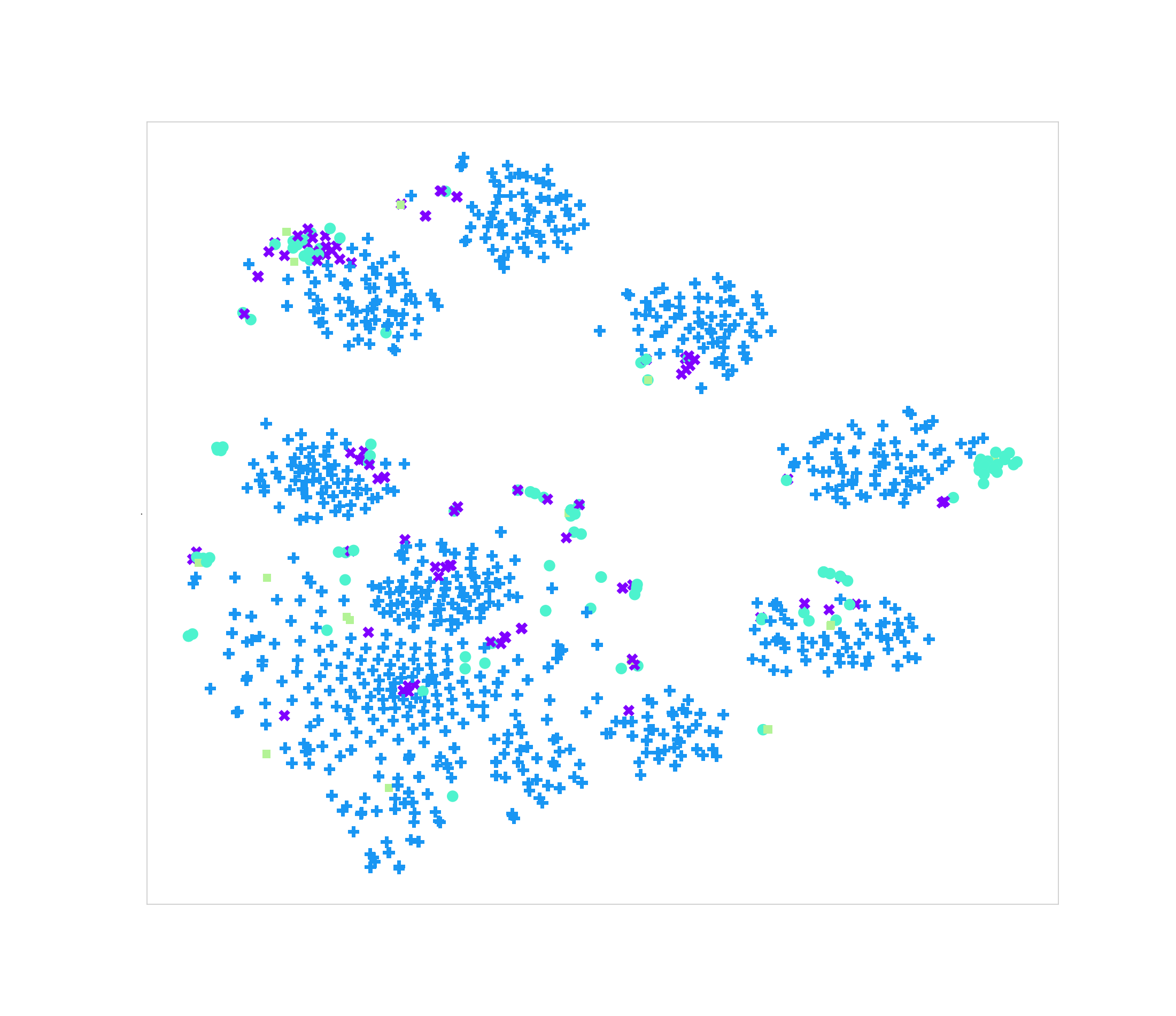}
         \caption{Seqvec}
         \label{tsne_seqvec_BC}
     \end{subfigure}
     \hfill
     \begin{subfigure}[b]{0.23\textwidth}
         \centering
         \includegraphics[scale=0.075]{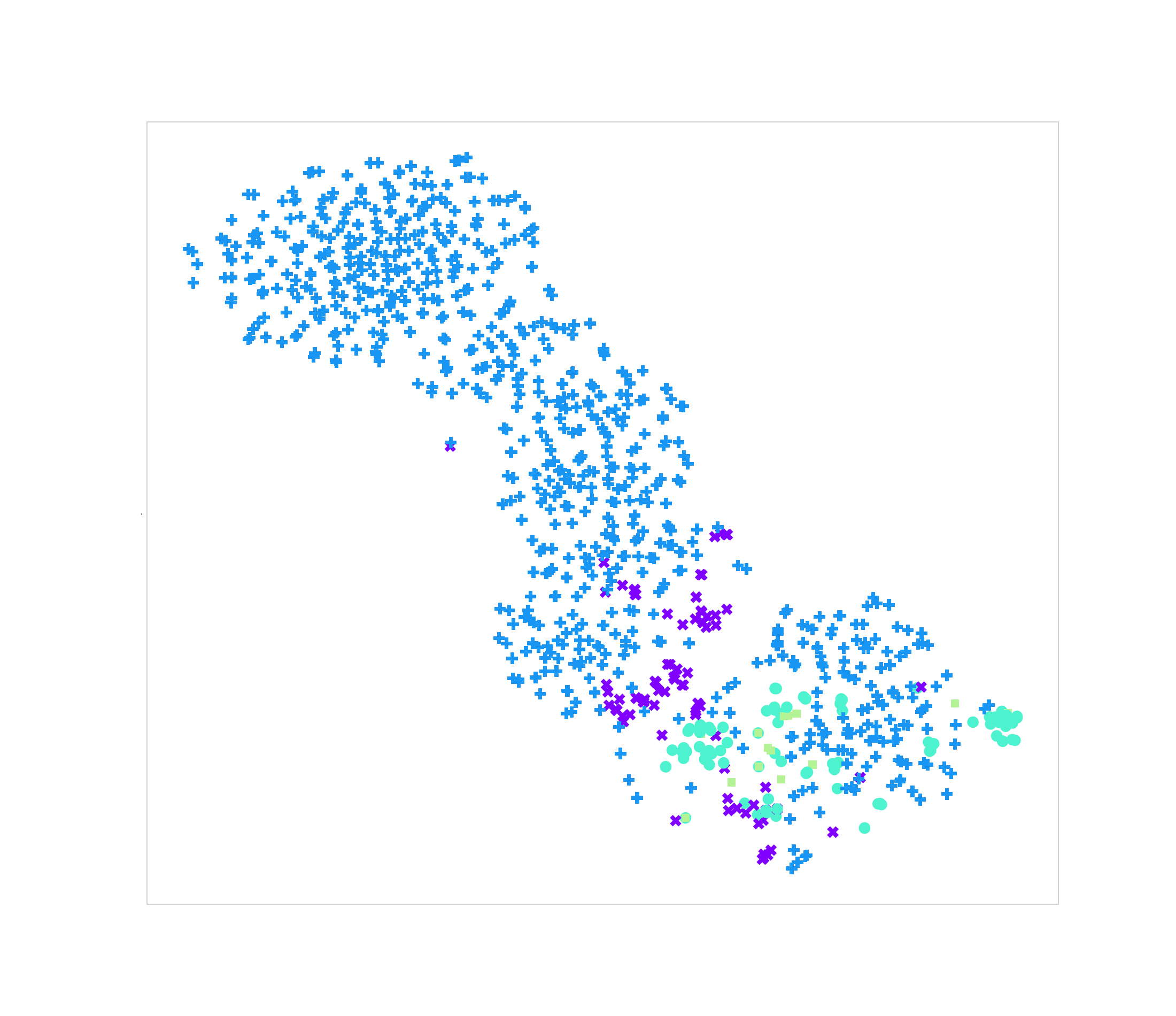}
         \caption{ESM2}
         \label{tsne_esm2_BC}
     \end{subfigure}
     \begin{subfigure}[b]{0.23\textwidth}
         \centering
         \includegraphics[scale=0.075]{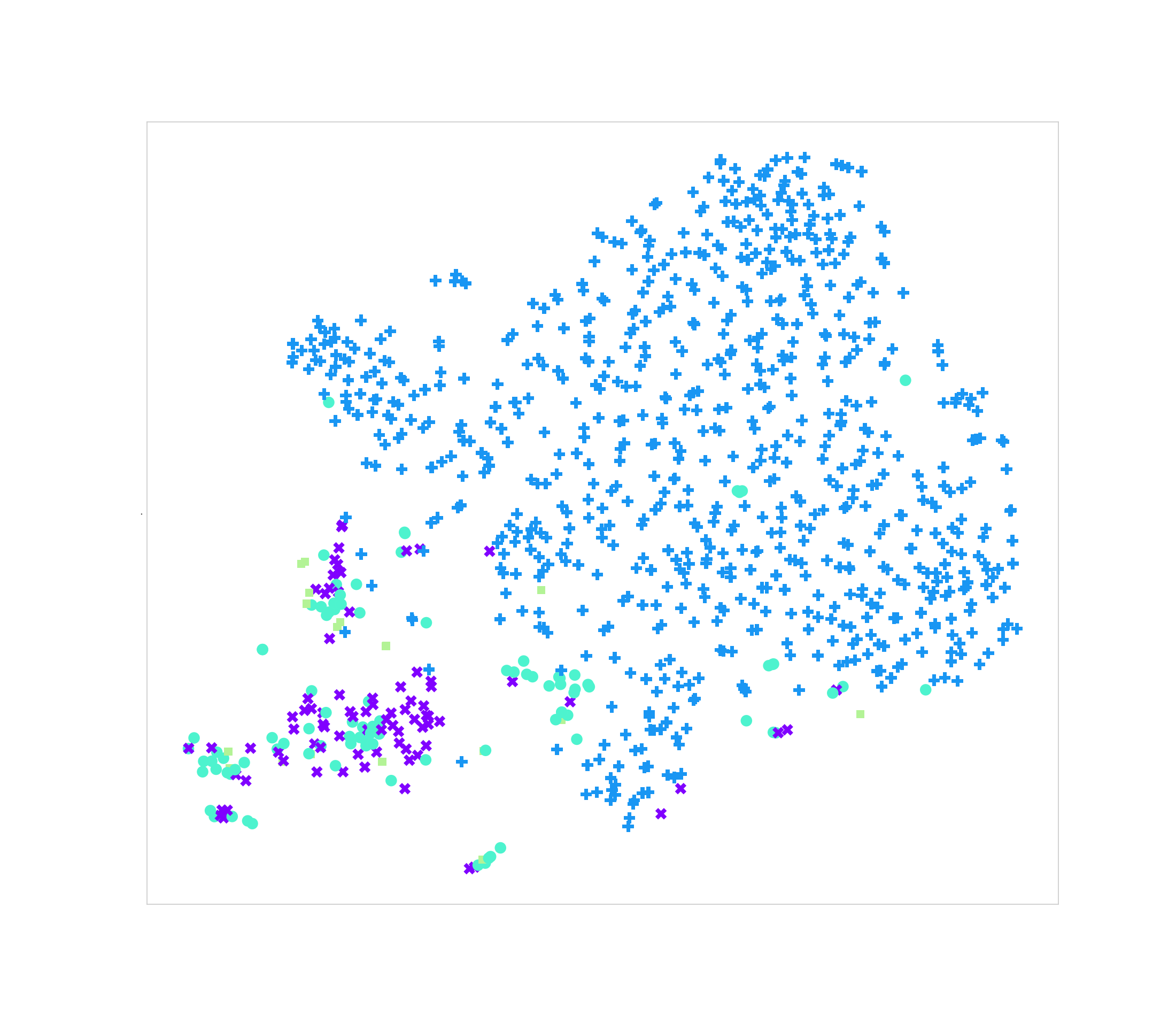}
         \caption{TAPE}
         \label{tsne_tape_BC}
     \end{subfigure}
          \hfill
     \begin{subfigure}[b]{0.23\textwidth}
         \centering
         \includegraphics[scale=0.075]{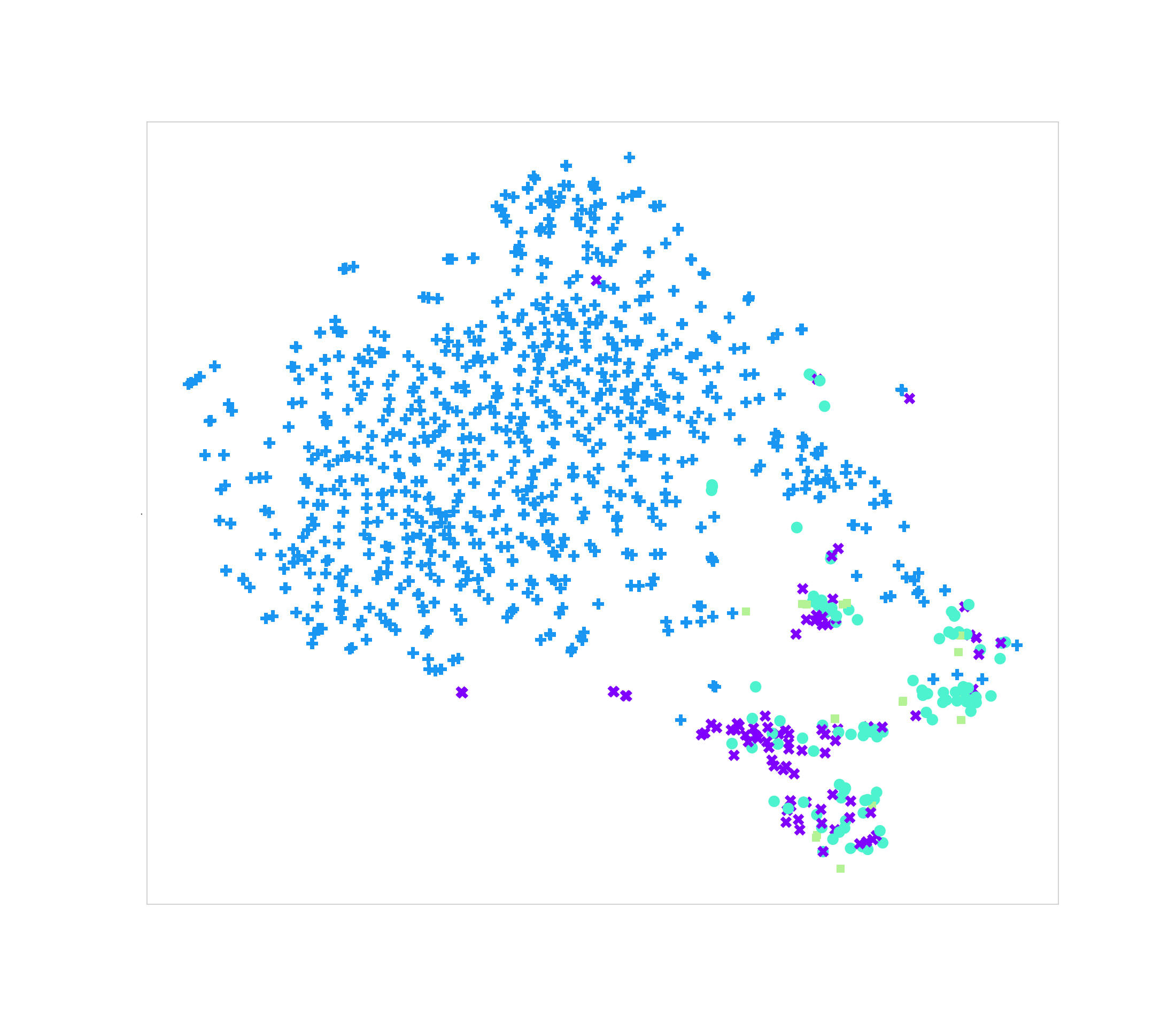}
         \caption{ProtT5}
         \label{tsne_protT5_BC}
     \end{subfigure}
     \includegraphics[width=0.75\textwidth]{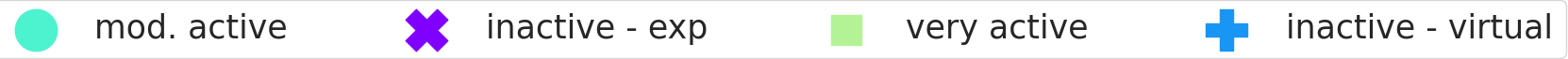}
     \caption{t-SNE visualization of different embeddings from\textbf{ Breast Cancer dataset}. The Figure is best seen in color.}
     \label{tsne_breast_cancer}
\end{figure*}

A popular visualization technique, named t-SNE~\citeauthor{van2008visualizing}, is employed to visualize the feature vectors generated by the SeqVec~\citeauthor{heinzinger2019modeling} method. 
The t-SNE plots are shown in Figure~\ref{tsne_pdb186}  for PDB186, Figure~\ref{tsne_lung_cancer} for Lung Cancer, and Figure~\ref{tsne_breast_cancer} for breast cancer datasets. The main idea for reporting the t-SNE plots is to assess whether there is a clear class separation between the labels in the datasets using embeddings generated by the respective LLMs. 
We observe that clusters overlap for Seqvec in the Lung cancer and breast cancer datasets, indicating that no clear decision boundary exists in the data.

\section{Results And Discussion}
\label{sec_results}

In the first step, we simulate the data to understand the hyperbolicity values for known structures using the following synthetic data.
\paragraph{Sphere Space Data Computation} 
To simulate data in the sphere space, we generate $n$ random points on the surface of a unit sphere in a high-dimensional space. Each point is sampled from a standard normal distribution in $\mathbb{R}^d$, and subsequently normalized to lie on the unit sphere by dividing each point by its Euclidean norm. The distance matrix for these points is computed using the Euclidean distance metric. The $\delta$-hyperbolicity values for this space are then calculated by evaluating the shortest path distances between all possible pairs of points, followed by applying the $\delta$-hyperbolicity calculation method described in Section~\ref{sec_PAa}.

\paragraph{Dense Graph Data Computation} 
A dense graph is generated using the Erdős-Rényi model, where $n$ nodes are connected with a high probability, $p = 0.8$. The shortest path distances are computed using the Floyd–Warshall algorithm. $\delta$-hyperbolicity is evaluated from this distance matrix, assigning a large finite value to any disconnected pairs to keep the computation valid.

\paragraph{Poincaré Space Data Computation} 
To simulate data in the Poincaré space, we sample $n$ random points in the two-dimensional Poincaré ball by drawing radial distances $r \in [0,1)$ and angles $\theta \in [0,2\pi)$, then converting to Cartesian coordinates via $x = r\cos\theta$ and $y = r\sin\theta$. The geodesic distances of pairs are calculated using the hyperbolic distance formula, which incorporates both the Euclidean separation and the distance of each point from the origin.



\begin{table}[h!]
    \centering
    \resizebox{0.6\textwidth}{!}{
    \begin{tabular}{cc}
    \toprule
         Metric Space & $\delta$ \\
         \midrule \midrule
         Sphere Space - $\delta$: & 0.32 $\pm$ 0.19 \\
        Dense Graph - $\delta$: & 0.31 $\pm$ 0.25 \\
        Poincaré Space - $\delta$: & 0.2867 $\pm$ 0.3328 \\
        \bottomrule
    \end{tabular}
    }
    \caption{Hyperbolicity $\delta$ values (average and standard deviation) for synthetically generated data.}
    \label{tbl_synthetic}
\end{table}

The differences in $\delta$-hyperbolicity across the three synthetic spaces reflect their underlying geometry as shown in Table~\ref{tbl_synthetic}. Sphere and dense graph data show moderately higher $\delta$ values of $0.32 \pm 0.19$ and $0.31 \pm 0.25$, respectively, due to curvature and high connectivity, both of which reduce tree-likeness. The Poincaré space, on the other hand, produces the lowest $\delta$ value of $0.2867 \pm 0.3328$, which is consistent with its inherently hyperbolic, tree-like structure~\citeauthor{nickel2017poincare}. This illustrates how a space's tree-like appearance is shaped by geometry.

\subsection{Hierarchical, Geometric and Topological Results}

\subsubsection{PDB186 dataset}
Table \ref{tab_PDB186_embedding_metrics} summarizes the geometric structure of the embeddings across all metrics. TAPE has the highest tree-likeness in the Poincaré space, with the lowest fractal dimension, hyperbolicity, ultrametricity, and NJ deviation. SeqVec shows the most negative Ricci curvature and the highest persistence entropy, indicating strong local hyperbolicity but weaker global hierarchy.ProtT5 exhibits significant deviations from tree structure and performs poorly in Poincaré space.
With the lowest $\delta_{\text{avg}}$, $U_{\text{avg}}$, and NJ values under Euclidean distances, ProtT5 becomes the most tree-like, and ESM performs the worst. In general, Poincaré distances yield far smaller and more stable hierarchy metrics, confirming that hyperbolic geometry captures tree structure more effectively than Euclidean space.

Overall, the results show that (i) ProtT5 works best when utilizing Euclidean distances; (ii) TAPE is the most hierarchical in hyperbolic space; and (iii) SeqVec shows great local hyperbolicity but weaker global hierarchy. Results show that the tree-likeness of the resulting representation space is significantly influenced by both the embedding model and the underlying distance geometry.

\begin{table}[ht]
\centering
\LARGE
\resizebox{0.9\textwidth}{!}{
\begin{tabular}{lccccccc}
\toprule
\multirow{1}{*}{Dist Matrix} & Embedding
& $\boldsymbol{\delta_{\text{avg}}}$ 
& $\boldsymbol{U_{\text{avg}}}$ 
& $\boldsymbol{\mathrm{NJ}_{\text{avg}}}$ 
& $\boldsymbol{\mathrm{Ricci}_{\text{avg}}}$ 
& $\boldsymbol{\mathrm{PersEnt}_{\text{avg}}}$ 
& $\boldsymbol{\mathrm{FracDim}_{\text{avg}}}$ \\
\midrule
\multirow{4}{*}{Poincare}
& SeqVec  & 0.0110 & 0.0242 & 30.5558 & \textbf{-0.1712} & \textbf{19.0043} & 29.9867	\\
& ESM     & 0.0183 & 0.0124 & 19.3127 & -0.0650 & 0.3970 & inf        \\
& TAPE    & \textbf{0.0030} & \textbf{0.0083} & \textbf{16.3208} & -0.0642 & 7.3302 & \textbf{16.3534}    \\
& ProtT5  & 0.0446 & 0.3038 & 567.0299 & -0.0652 & 1.1681 & 19.4370     \\
\midrule
\multirow{4}{*}{Euclidean}
& SeqVec  & 1.6227 & 3.6570 & 7276.7976 & -0.1712 & 19.0043  & 29.9867	\\
& ESM     & 0.8727 & 0.7193 & 920.0283 & -0.0650 & 0.3970 & inf           \\
& TAPE    & 0.4178 & 1.0126 & 2123.9591 & -0.0642 & 7.3302 & 16.3534      \\
& ProtT5  & \textbf{0.0418} & \textbf{0.1300} & \textbf{318.6305} & -0.0652 & 1.1681 & 19.4370       \\
\bottomrule
\end{tabular}
}
\caption{Comparison of geometric and topological metrics across different embeddings for the PDB186 dataset, grouped by the distance matrix. The log ratio in the fractal dimension formula approaches zero, and its reciprocal explodes to infinity as a result of ESM embeddings placing numerous nearest neighbors at approximately identical distances, which results in the inf value.}
\label{tab_PDB186_embedding_metrics}
\end{table}

\subsubsection{Lung Cancer dataset}

The Lung Cancer results show that geometric behavior varies strongly across embeddings and distance metrics. As shown in Table~\ref{tab_lung_cancer_embedding_metrics}, in the Poincaré space, ESM2 is the most hierarchical, with the lowest hyperbolicity (0.0016), ultrametricity (0.0022), and NJ deviation (40.29), as well as the most negative Ricci curvature. TAPE shows the lowest fractal dimension (18.57), indicating a more compact manifold, while SeqVec exhibits the highest persistence entropy, reflecting greater multiscale variability. ProtT5 shows weak hierarchy in Poincaré space but becomes the most tree-like under Euclidean distances, with the lowest $\delta$ (0.1010), $U$ (0.2413), and $NJ$ deviation (2296.54). Overall, ESM2 and TAPE demonstrate stronger hyperbolic structure, whereas ProtT5 appears more separable in Euclidean geometry.

\begin{table}[ht]
\centering
\LARGE
\resizebox{0.9\textwidth}{!}{
\begin{tabular}{lccccccc}
\toprule
\multirow{1}{*}{Dist Matrix} & Embedding
& $\boldsymbol{\delta_{\text{avg}}}$ 
& $\boldsymbol{U_{\text{avg}}}$ 
& $\boldsymbol{\mathrm{NJ}_{\text{avg}}}$ 
& $\boldsymbol{\mathrm{Ricci}_{\text{avg}}}$ 
& $\boldsymbol{\mathrm{PersEnt}_{\text{avg}}}$ 
& $\boldsymbol{\mathrm{FracDim}_{\text{avg}}}$ \\
\midrule
\multirow{4}{*}{Poincare}
& SeqVec & 0.0026 & 0.0069 & 118.3239 	& -0.0637 & \textbf{22.4044} & 32.0288	\\
& ESM &  \textbf{0.0016} & \textbf{0.0022} &  \textbf{40.2858} & \textbf{-0.0849} & 20.7693 & 32.9002 \\
& TAPE & 0.0040  & 0.0084 &  72.4654   & -0.0653 & 6.4817 &  \textbf{18.5660} \\
& ProtT5 & 0.0305 & 0.0759 & 715.5995  & -0.0621 & 1.4096 &  20.9063 \\

\midrule
\multirow{4}{*}{Euclidean}
& SeqVec & 0.5904 & 1.2987 & 28674.5853 & -0.0637 & \textbf{22.4044} & 32.0288 \\
& ESM    & 1.0897 & 1.4986 & 27513.1399 &\textbf{ -0.0849} & 20.7693 & 32.9002 \\
& TAPE   & 0.5944 & 1.2251 & 10738.2480 & -0.0653 & 6.4817 &  \textbf{18.5660} \\
& ProtT5 & \textbf{0.1010} & \textbf{0.2413} & \textbf{2296.5416} & -0.0621 & 1.4096 &  20.9063 \\
\bottomrule
\end{tabular}
}
\caption{Comparison of geometric and topological metrics across different embeddings for the Lung Cancer dataset, grouped by the distance matrix.}
\label{tab_lung_cancer_embedding_metrics}
\end{table}

\subsubsection{Breast Cancer dataset}
The Breast Cancer results in Table~\ref{tab_breast_cancer_embedding_metrics} show consistent geometric differences across embeddings. In the Poincaré space, ESM2 is the most hierarchical, achieving the lowest hyperbolicity (0.0017), ultrametricity (0.0023), and NJ deviation (42.74), along with the most negative Ricci curvature. Here also, TAPE shows the lowest fractal dimension (18.12), indicating a compact embedding manifold, while SeqVec exhibits the highest persistence entropy (22.23), reflecting more topological diversity. 
With significantly higher ultrametricity (0.0758) and NJ values (752.72), ProtT5 exhibits the weakest Poincaré hierarchy. Similar to patterns seen in previous datasets, ProtT5 becomes the most tree-like under Euclidean distances, exhibiting the lowest $\delta$ (0.1073), $U$ (0.2458), and $NJ$ deviation (2487.96). Overall, we can conclude here that ESM2 and TAPE exhibit stronger hyperbolic structure, whereas ProtT5 appears more separable in Euclidean space, and SeqVec remains more topologically fragmented.

\begin{table}[ht]
\centering
\LARGE
\resizebox{0.9\textwidth}{!}{
\begin{tabular}{lccccccc}
\toprule
\multirow{1}{*}{Dist Matrix} & Embedding
& $\boldsymbol{\delta_{\text{avg}}}$ 
& $\boldsymbol{U_{\text{avg}}}$ 
& $\boldsymbol{\mathrm{NJ}_{\text{avg}}}$ 
& $\boldsymbol{\mathrm{Ricci}_{\text{avg}}}$ 
& $\boldsymbol{\mathrm{PersEnt}_{\text{avg}}}$ 
& $\boldsymbol{\mathrm{FracDim}_{\text{avg}}}$ \\
\midrule
\multirow{4}{*}{Poincare}
& SeqVec &  0.0025 & 0.0073 & 124.5329  & -0.0675 & 22.2295 & 30.0684 \\
& ESM    &  \textbf{0.0017} & \textbf{0.0023} & \textbf{42.7396} & -0.0748 & 20.4887 & 31.0122 \\
& TAPE   &  0.0040 & 0.0087 & 76.5285  & -0.0690 & 6.4066 &  18.1227 \\
& ProtT5 &  0.0306 & 0.0758 & 752.7241 & -0.0653 & 1.4063 &  20.4609 \\

\midrule
\multirow{4}{*}{Euclidean}
& SeqVec & 0.5804 & 1.3425 & 30408.3304 & -0.0675 & \textbf{22.2295} & 30.0684 \\
& ESM    & 1.1336 & 1.5309 & 28876.8367 & \textbf{-0.0748} & 20.4887 & 31.0122 \\
& TAPE   & 0.6024 & 1.2547 & 11401.2842 & -0.0690 & 6.4066 &  \textbf{18.1227} \\
& ProtT5 & \textbf{0.1073} & \textbf{0.2458} & \textbf{2487.9561} & -0.0653 & 1.4063 &  20.4609 \\
\bottomrule
\end{tabular}
}
\caption{Comparison of geometric and topological metrics across different embeddings for the Breast Cancer dataset, grouped by the distance matrix.}
\label{tab_breast_cancer_embedding_metrics}
\end{table}

\begin{remark}
    Because Ricci curvature, persistent homology, and fractal dimension are calculated directly on the embeddings rather than the complete pairwise distance matrix, they remain constant across the Euclidean and Poincaré distance matrices.
\end{remark}

\subsection{Classification Results}


The classification results on the PDB186 dataset are shown in Table~\ref{tbl_all_avg_data_results_pdb186}. ProtT5 achieves the highest ROC-AUC (0.7968 with LR), followed by ESM2 (0.7453), while SeqVec shows poor performance (0.6173). We can see that across other metrics as well, ProtT5 consistently outperforms SeqVec.
\begin{table}[h!]
      \centering
      \scriptsize
      \resizebox{0.9\textwidth}{!}{
\begin{tabular}{@{\extracolsep{6pt}}p{0.5cm}lp{0.5cm}p{0.5cm}p{0.6cm}p{0.7cm}p{0.7cm}p{0.5cm}p{1.5cm}}
    \toprule
        \multirow{2}{*}{Embed.} & \multirow{2}{*}{Algo.} & \multirow{2}{*}{Acc. $\uparrow$} & \multirow{2}{*}{Prec. $\uparrow$} & \multirow{2}{*}{Recall $\uparrow$} & \multirow{2}{1.2cm}{F1 (Weig.) $\uparrow$} & \multirow{2}{1.2cm}{F1 (Macro) $\uparrow$} & \multirow{2}{1cm}{ROC AUC $\uparrow$} & Train Time (sec.) $\downarrow$\\
        \midrule \midrule
      \multirow{7}{1.2cm}{SeqVec}
& SVM & 0.5786 & 0.6045 & 0.5786 & 0.5720 & 0.5755 & 0.5917 & 1692.2908		\\
& NB & 0.5036 & 0.5538 & 0.5036 & 0.3878 & 0.3836 & 0.5014 & \underline{132.7588}       \\
& MLP & 0.5607 & 0.5934 & 0.5607 & 0.5088 & 0.4911 & 0.5330 & 845.1396      \\
& KNN & 0.5107 & 0.3741 & 0.5107 & 0.3815 & 0.3812 & 0.5135 & 1808.8480     \\
& RF & 0.5857 & 0.5961 & 0.5857 & 0.5857 & 0.5856 & 0.5908 & 100.2860       \\
& LR & \underline{0.6179} & \underline{0.6216} & \underline{0.6179} & \underline{0.6184} & \underline{0.6159} & \underline{0.6173} & 1024.6331      \\
& DT & 0.5536 & 0.5550 & 0.5536 & 0.5531 & 0.5494 & 0.5507 & 277.4930       \\
 
\midrule
\multirow{7}{1.2cm}{ESM2}
& SVM & 0.7321 & 0.7520 & 0.7321 & 0.7327 & 0.7319 & 0.7420 & 0.0138	\\
& NB & 0.6321 & 0.6359 & 0.6321 & 0.6325 & 0.6273 & 0.6289 & \underline{0.0034}     \\
& MLP & 0.7107 & 0.7274 & 0.7107 & 0.7116 & 0.7103 & 0.7190 & 0.7072    \\
& KNN & 0.6214 & 0.6531 & 0.6214 & 0.6159 & 0.6178 & 0.6354 & 0.0171    \\
& RF & 0.6964 & 0.7231 & 0.6964 & 0.6952 & 0.6951 & 0.7080 & 0.3342     \\
& LR & \underline{0.7357} & \underline{0.7561} & \underline{0.7357} & \underline{0.7361} & \underline{0.7353} & \underline{0.7453} & 0.0452     \\
& DT & 0.5857 & 0.5986 & 0.5857 & 0.5851 & 0.5821 & 0.5894 & 0.0674     \\

\midrule
\multirow{7}{1.2cm}{TAPE}
& SVM & 0.6286 & 0.6411 & 0.6286 & 0.6301 & 0.6269 & 0.6317 & 0.0185	\\
& NB & 0.6893 & 0.6965 & 0.6893 & 0.6897 & 0.6858 & 0.6896 & \underline{\textbf{0.0031}}     \\
& MLP & \underline{0.6964} & \underline{0.7170} & \underline{0.6964} & \underline{0.6950} & \underline{0.6939} & \underline{0.7032} & 33.7174   \\
& KNN & 0.6000 & 0.6134 & 0.6000 & 0.5991 & 0.5954 & 0.6023 & 0.5691    \\
& RF & 0.6714 & 0.6974 & 0.6714 & 0.6696 & 0.6684 & 0.6795 & 0.4725     \\
& LR & 0.6286 & 0.6398 & 0.6286 & 0.6301 & 0.6268 & 0.6315 & 0.1313     \\
& DT & 0.5179 & 0.5372 & 0.5179 & 0.5153 & 0.5149 & 0.5259 & 0.0619     \\

\midrule

\multirow{7}{1.2cm}{ProtT5}
& SVM & 0.7679 & 0.7958 & 0.7679 & 0.7681 & 0.7675 & 0.7813 & 0.0235	\\
& NB & 0.7250 & 0.7385 & 0.7250 & 0.7228 & 0.7181 & 0.7261 & \underline{0.0035}     \\
& MLP & 0.7464 & 0.7558 & 0.7464 & 0.7459 & 0.7415 & 0.7465 & 38.2089   \\
& KNN & 0.6821 & 0.6928 & 0.6821 & 0.6818 & 0.6784 & 0.6843 & 0.6559    \\
& RF & 0.7036 & 0.7291 & 0.7036 & 0.7029 & 0.7015 & 0.7136 & 0.3849     \\
& LR & \underline{\textbf{0.7893}} & \underline{\textbf{0.8029}} & \underline{\textbf{0.7893}} & \underline{\textbf{0.7903}} & \underline{\textbf{0.7884}} & \underline{\textbf{0.7968}} & 0.1281     \\
& DT & 0.5607 & 0.5826 & 0.5607 & 0.5585 & 0.5567 & 0.5687 & 0.0841     \\
     
    \bottomrule
  \end{tabular}  
    }
    \caption{Average classification results for different models and algorithms for \textbf{PDB186} dataset.}
    \label{tbl_all_avg_data_results_pdb186}
\end{table}
Table~\ref{tbl_all_avg_data_results_LC} shows the classification results for the Lung Cancer dataset. Here, ESM2 achieves the highest ROC-AUC (0.8568 with SVM), followed by ProtT5 (0.8311 with LR). SeqVec and TAPE consistently show the lowest performance, with only marginal differences between them across other metrics.
\begin{table}[h!]
      \centering
      \scriptsize
      \resizebox{0.9\textwidth}{!}{
\begin{tabular}{@{\extracolsep{6pt}}p{0.5cm}lp{0.5cm}p{0.5cm}p{0.6cm}p{0.7cm}p{0.7cm}p{0.5cm}p{1.5cm}}
    \toprule
        \multirow{2}{*}{Embed.} & \multirow{2}{*}{Algo.} & \multirow{2}{*}{Acc. $\uparrow$} & \multirow{2}{*}{Prec. $\uparrow$} & \multirow{2}{*}{Recall $\uparrow$} & \multirow{2}{1.2cm}{F1 (Weig.) $\uparrow$} & \multirow{2}{1.2cm}{F1 (Macro) $\uparrow$} & \multirow{2}{1cm}{ROC AUC $\uparrow$} & Train Time (sec.) $\downarrow$\\
        \midrule \midrule

      \multirow{7}{1.2cm}{SeqVec}
& SVM & \underline{0.9240} & \underline{0.9212} & \underline{0.9240} & \underline{0.9202} & \underline{0.6836} & \underline{0.8212} & 29.3541		\\
& NB & 0.6347 & 0.7938 & 0.6347 & 0.6212 & 0.2585 & 0.6396 & \underline{3.6623}         \\
& MLP & 0.9048 & 0.9101 & 0.9048 & 0.9056 & 0.6513 & 0.8121 & 142.6836      \\
& KNN & 0.8812 & 0.8660 & 0.8812 & 0.8683 & 0.5436 & 0.7165 & 9.0200        \\
& RF & 0.9085 & 0.8955 & 0.9085 & 0.8977 & 0.6329 & 0.7649 & 7.2720         \\
& LR & 0.7882 & 0.8925 & 0.7882 & 0.8254 & 0.5433 & 0.8066 & 1005.2145      \\
& DT & 0.8531 & 0.8692 & 0.8531 & 0.8593 & 0.5348 & 0.7469 & 23.7610        \\

\midrule
\multirow{7}{1.2cm}{ESM2}
& SVM & \underline{\textbf{0.9506}} & \underline{\textbf{0.9474}} & \underline{\textbf{0.9506}} & \underline{\textbf{0.9475}} & \underline{\textbf{0.7335}} & \underline{\textbf{0.8568}} & 11.9609		\\
& NB & 0.6472 & 0.8906 & 0.6472 & 0.7109 & 0.5231 & 0.8066 & \underline{0.6002}         \\
& MLP & 0.9365 & 0.9453 & 0.9365 & 0.9389 & 0.7428 & 0.8698 & 9.4761        \\
& KNN & 0.9122 & 0.9259 & 0.9122 & 0.9144 & 0.6687 & 0.8300 & 1.2228        \\
& RF & 0.9380 & 0.9352 & 0.9380 & 0.9342 & 0.7144 & 0.8296 & 2.8020         \\
& LR & 0.7823 & 0.8957 & 0.7823 & 0.8171 & 0.5472 & 0.8243 & 156.7906       \\
& DT & 0.9011 & 0.9152 & 0.9011 & 0.9039 & 0.6469 & 0.8074 & 5.3667         \\

\midrule
\multirow{7}{1.2cm}{TAPE}
& SVM & 0.9166 & 0.9181 & 0.9166 & 0.9165 & 0.6177 & 0.7967 & 0.0394	\\
& NB & 0.9018 & 0.9204 & 0.9018 & 0.9082 & 0.6461 & \underline{0.8306} & \underline{\textbf{0.0095}}     \\
& MLP & 0.9085 & 0.9152 & 0.9085 & 0.9102 & 0.6194 & 0.7943 & 1.7274    \\
& KNN & 0.9122 & 0.9143 & 0.9122 & 0.9107 & 0.5994 & 0.7771 & 0.0265    \\
& RF & 0.9144 & 0.9065 & 0.9144 & 0.9054 & 0.6336 & 0.7546 & 0.9389     \\
& LR & \underline{0.9173} & \underline{0.9252} & \underline{0.9173} & \underline{0.9200} & \underline{0.6564} & 0.8221 & 0.7910     \\
& DT & 0.8657 & 0.8873 & 0.8657 & 0.8748 & 0.5586 & 0.7576 & 0.3775     \\

\midrule
\multirow{7}{1.2cm}{ProtT5}
& SVM & 0.9277 & 0.9313 & 0.9277 & 0.9277 & 0.6664 & 0.8290 & 0.0513	\\
& NB & 0.9151 & 0.9226 & 0.9151 & 0.9175 & 0.6461 & 0.8232 & \underline{0.0125}     \\
& MLP & 0.9196 & 0.9263 & 0.9196 & 0.9219 & 0.6375 & 0.8155 & 3.5043    \\
& KNN & 0.9188 & 0.9169 & 0.9188 & 0.9165 & 0.5826 & 0.7817 & 0.0649    \\
& RF & \underline{0.9321} & 0.9293 & \underline{0.9321} & 0.9258 & 0.6520 & 0.7997 & 1.0917     \\
& LR & 0.9277 & \underline{0.9326} & 0.9277 & \underline{0.9288} & \underline{0.6724} & \underline{0.8311} & 1.0877     \\
& DT & 0.8841 & 0.8977 & 0.8841 & 0.8887 & 0.5800 & 0.7800 & 0.4553     \\
    \bottomrule
  \end{tabular}  
    }
    \caption{Average classification results for different models and algorithms for \textbf{Lung Cancer} dataset.}
    \label{tbl_all_avg_data_results_LC}
\end{table}
The classification results for the Breast Cancer dataset are shown in Table~\ref{tbl_all_avg_data_results_BC}, and they reinforce the trend observed previously. ESM2 attains the highest ROC-AUC (0.8385), followed by ProtT5 at 0.8195, while SeqVec again shows the weakest performance (0.7597). Across all other metrics, SeqVec consistently lags behind by a noticeable margin, further highlighting the superiority of transformer-based embeddings for classification tasks.

\begin{table}[h!]
      \centering
      \scriptsize
      \resizebox{0.9\textwidth}{!}{
\begin{tabular}{@{\extracolsep{6pt}}p{0.5cm}lp{0.5cm}p{0.5cm}p{0.6cm}p{0.7cm}p{0.7cm}p{0.5cm}p{1.5cm}}
    \toprule
        \multirow{2}{*}{Embed.} & \multirow{2}{*}{Algo.} & \multirow{2}{*}{Acc. $\uparrow$} & \multirow{2}{*}{Prec. $\uparrow$} & \multirow{2}{*}{Recall $\uparrow$} & \multirow{2}{1.2cm}{F1 (Weig.) $\uparrow$} & \multirow{2}{1.2cm}{F1 (Macro) $\uparrow$} & \multirow{2}{1cm}{ROC AUC $\uparrow$} & Train Time (sec.) $\downarrow$\\
        \midrule \midrule
      \multirow{7}{1.2cm}{SeqVec}
& SVM & \underline{0.9004} & \underline{0.8995} & \underline{0.9004} & \underline{0.8977} & \underline{0.5682} & \underline{0.7597} & 1724.5	\\
& NB & 0.4996 & 0.7564 & 0.4996 & 0.5289 & 0.2087 & 0.5882 & \underline{4.2771} \\
& MLP & 0.8716 & 0.8833 & 0.8716 & 0.8740 & 0.5218 & 0.7447 & 277.84 \\
& KNN & 0.7761 & 0.8375 & 0.7761 & 0.7969 & 0.4523 & 0.6859 & 175.88 \\
& RF & 0.8814 & 0.8649 & 0.8814 & 0.8671 & 0.5349 & 0.7082 & 11.531 \\
& LR & 0.7719 & 0.8811 & 0.7719 & 0.8164 & 0.5014 & 0.7528 & 1836.9 \\
& DT & 0.8133 & 0.8245 & 0.8133 & 0.8179 & 0.4515 & 0.6745 & 37.130 \\

\midrule
\multirow{7}{1.2cm}{ESM2} 
& SVM & \underline{\textbf{0.9333}} & \underline{\textbf{0.9335}} & \underline{\textbf{0.9333}} & \underline{\textbf{0.9319}} & \underline{\textbf{0.6843}} & \underline{\textbf{0.8385}} & 8.0102 \\
& NB & 0.6681 & 0.8646 & 0.6681 & 0.7150 & 0.5013 & 0.7977 & \underline{0.7043} \\
& MLP & 0.9067 & 0.9182 & 0.9067 & 0.9105 & 0.6353 & 0.8032 & 12.827 \\
& KNN & 0.9060 & 0.9089 & 0.9060 & 0.9047 & 0.6380 & 0.8268 & 1.1774 \\
& RF & 0.9326 & 0.9257 & 0.9326 & 0.9261 & 0.6589 & 0.7914 & 2.7038 \\
& LR & 0.7993 & 0.8796 & 0.7993 & 0.8239 & 0.5591 & 0.8177 & 154.53 \\
& DT & 0.8877 & 0.8889 & 0.8877 & 0.8871 & 0.6237 & 0.7948 & 3.7809 \\

\midrule
\multirow{7}{1.2cm}{TAPE}
& SVM & 0.8982 & 0.8968 & 0.8982 & 0.8962 & 0.5964 & 0.7813 & 0.0930	\\
& NB & 0.8575 & 0.8852 & 0.8575 & 0.8691 & 0.5760 & 0.7887 & \underline{\textbf{0.0122}}     \\
& MLP & 0.8884 & 0.8943 & 0.8884 & 0.8901 & 0.5825 & 0.7815 & 284.43  \\
& KNN & 0.8821 & 0.8716 & 0.8821 & 0.8745 & 0.5119 & 0.7345 & 1.1070    \\
& RF & 0.8870 & 0.8683 & 0.8870 & 0.8738 & 0.5615 & 0.7374 & 2.2642     \\
& LR & \underline{0.9046} & \underline{0.9082} & \underline{0.9046} & \underline{0.9030} & \underline{0.6161} & \underline{0.7961} & 1.9700     \\
& DT & 0.8386 & 0.8511 & 0.8386 & 0.8433 & 0.5123 & 0.7262 & 0.9652     \\

\midrule
\multirow{7}{1.2cm}{ProtT5}
& SVM & 0.8975 & 0.9011 & 0.8975 & 0.8973 & 0.6425 & 0.8086 & 0.0743	\\
& NB & 0.8842 & 0.8902 & 0.8842 & 0.8865 & 0.5744 & 0.7770 & \underline{0.0142}     \\
& MLP & 0.8961 & \underline{0.9015} & 0.8961 & \underline{0.8977} & \underline{0.6509} & \underline{0.8195} & 3.3378    \\
& KNN & 0.8933 & 0.8796 & 0.8933 & 0.8855 & 0.5230 & 0.7425 & 0.0496    \\
& RF & \underline{0.9018} & 0.8892 & \underline{0.9018} & 0.8934 & 0.5996 & 0.7651 & 1.3799     \\
& LR & 0.8947 & 0.8973 & 0.8947 & 0.8951 & 0.6131 & 0.8004 & 1.0668     \\
& DT & 0.8484 & 0.8535 & 0.8484 & 0.8497 & 0.5021 & 0.7278 & 0.6561     \\
    \bottomrule
  \end{tabular}  
    }
    \caption{Average classification results for different models and algorithms for \textbf{Breast Cancer} dataset.}
    \label{tbl_all_avg_data_results_BC}
\end{table}

\subsection{Clustering Results}
Table~\ref{tbl_clustering_PDB_186} presents the clustering results for the PDB186 dataset, while Table~\ref{tbl_clustering_Lung_Cancer} and Table~\ref{tbl_clustering_Breast_Cancer} show the results for the Lung Cancer and Breast Cancer datasets, respectively.

For the PDB186 dataset, SeqVec shows an interesting contrast: it performs best on clustering metrics such as the Silhouette Coefficient and Davies–Bouldin Score in all clustering algorithms, yet it is the weakest model in the classification setting. ESM2, which ranked second in classification, shows more consistent behavior, particularly on the Calinski–Harabasz Score, where it performs well for Agglomerative and k-means clustering. This contrast highlights how models that excel in supervised classification do not necessarily produce the most cluster-friendly embedding spaces, and vice versa.


For the Lung Cancer dataset, ProtT5 performs better in all clustering metrics for Agglomerative and k-means. However, the k-modes findings are more in line with classification: ESM2, which is marginally more effective in supervised tasks, also receives higher k-modes scores, particularly on the Silhouette and Calinski–Harabasz scores. We can say that ProtT5 yields more cluster-friendly continuous embeddings, whereas ESM2 captures a structure better suited to categorical clustering.


We find a trend comparable to the Lung Cancer results in the Breast Cancer dataset. Although ESM2 outperforms ProtT5 in classification, it performs better in k-modes, particularly on the Silhouette and Calinski–Harabasz scores. ProtT5 has the best clustering performance for Agglomerative and k-means. This again exhibits a similar pattern: ProtT5 produces more cluster-friendly continuous embeddings, while ESM2 captures structure more closely aligned with categorical clustering and supervised classification tasks.

\begin{table}[h!]
  \centering
  \scriptsize
  \resizebox{0.9\textwidth}{!}{
  \begin{tabular}{p{1.8cm}p{1.2cm}p{1.2cm}p{1.4cm}p{1.4cm}p{1.3cm}p{1.8cm}}
    \toprule
    & & \multicolumn{3}{c}{Evaluation Metrics} \\
    \cmidrule{3-5}
    Algorithm & Embedding & Silhouette Coefficient $\uparrow$ & Calinski-Harabasz Score $\uparrow$ & Davies-Bouldin Score $\downarrow$ & Clustering Runtime $\downarrow$\\
    \midrule \midrule
    \multirow{4}{*}{Agglomerative} & 
     Seqvec & \textbf{\underline{0.603}} & 9.872 & \textbf{\underline{0.278}} &  \textbf{\underline{0.01}} \\
    & ESM2 & 0.230 & \underline{44.368} & 1.348 &  0.02 \\
    & TAPE & 0.136 & 12.068 & 3.665 & \textbf{\underline{0.01}} \\
    & ProtT5 & 0.117 & 10.429 &  2.306 & 0.03 \\
    \midrule
    \multirow{4}{*}{$k$-means} & 
     Seqvec &  \underline{0.384} & 3.407 & \underline{0.469} &  1.46 \\
    & ESM2 & 0.178 & \textbf{\underline{51.234}} & 1.820 &  1.46 \\
    & TAPE & 0.096 & 13.810 & 3.567 & \underline{1.20} \\
    & ProtT5 & 0.055 & 11.117 &  3.963 & 1.78 \\
    \midrule
    \multirow{4}{*}{$k$-modes} & 
     Seqvec & \underline{0.312} & \underline{2.545} & \underline{0.543} &  \underline{0.73} \\
    & ESM2 & 0.161 & 2.118 & 0.657 &  3.58 \\
    & TAPE & 0.033 & 1.167 & 0.895 &  2.32 \\
    & ProtT5 & -0.075 & 0.698 & 1.161 &  2.73 \\
    \bottomrule
  \end{tabular}
  }
  \caption{Internal clustering quality metrics for $k$-means
    and $k$-modes on different embeddings on the
    PDB186. Best values are shown in bold.}
  \label{tbl_clustering_PDB_186}
\end{table}

\begin{table}[h!]
  \centering
    \scriptsize
  \resizebox{0.9\textwidth}{!}{
  \begin{tabular}{p{1.8cm}p{1.2cm}p{1.2cm}p{1.4cm}p{1.4cm}p{1.3cm}}
    \toprule
    & & \multicolumn{3}{c}{Evaluation Metrics} \\
    \cmidrule{3-5}
    Algorithm & Embedding & Silhouette Coefficient $\uparrow$ & Calinski-Harabasz Score $\uparrow$ & Davies-Bouldin Score $\downarrow$ & Clustering Runtime $\downarrow$\\
    \midrule \midrule
     \multirow{4}{*}{Agglomerative} & 
     Seqvec & -0.002 & 36.713 & 3.792 & \textbf{\underline{0.22}} \\
    & ESM2 & 0.048 & 75.235 & 3.012 &  9.67 \\
    & TAPE & 0.082 & 80.764 & 2.709 &  0.23 \\
    & ProtT5 & \underline{0.177} & \underline{88.294} & \underline{2.030} &  0.25 \\
    \midrule
    \multirow{4}{*}{$k$-means} & 
     Seqvec & 0.026 & 38.255 & 3.922 & 2.29 \\
    & ESM2 & 0.070 & 79.774 & 3.370 &  5.95 \\
    & TAPE & 0.118 & \textbf{\underline{98.614}} & 2.527 &  2.54 \\
    & ProtT5 & \textbf{\underline{0.197}} & 94.129 & \underline{1.947} &  \underline{1.58} \\
    \midrule
    \multirow{3}{*}{$k$-modes} & 
     Seqvec & -0.038 & 0.947 & \underline{1.023} &  \underline{16.12} \\
    & ESM2 & \underline{0.035} & \underline{44.883} & 6.223 &  1642.4 \\
    & TAPE &  -0.168 & 0.837 & 1.087 &  18.40 \\
    & ProtT5 & -0.162 & 1.388 & \textbf{\underline{1.023}} & 22.34 \\
    \bottomrule
  \end{tabular}
  }
  \caption{Internal clustering quality metrics for $k$-means
    and $k$-modes on different embeddings on the
    Lung Cancer. Best values are shown in bold.}
  \label{tbl_clustering_Lung_Cancer}
\end{table}

\begin{table}[h!]
  \centering
    \scriptsize
  \resizebox{0.9\textwidth}{!}{
 \begin{tabular}{p{1.8cm}p{1.2cm}p{1.2cm}p{1.4cm}p{1.4cm}p{1.3cm}}
    \toprule
    & & \multicolumn{3}{c}{Evaluation Metrics} \\
    \cmidrule{3-5}
    Algorithm & Embedding & Silhouette Coefficient $\uparrow$ & Calinski-Harabasz Score $\uparrow$ & Davies-Bouldin Score $\downarrow$ & Clustering Runtime $\downarrow$\\
    \midrule	\midrule	
     \multirow{4}{*}{Agglomerative} & 
     Seqvec & -0.002 & 38.110 & 3.868 & 0.30 \\
    & ESM2 & 0.048 & 78.754 & 3.064 & 10.75 \\
    & TAPE & 0.096 & 87.078 & 2.826 & 0.17 \\
    & ProtT5 & \textbf{\underline{0.206}} & \underline{100.927} & \underline{2.003} & \textbf{\underline{0.14}} \\
    \midrule
    \multirow{4}{*}{$k$-means} & 
     Seqvec & -0.004 & 37.709 & 3.765 &  2.41 \\
    & ESM2 & 0.072 & 84.914 & 3.380 &  5.71 \\
    & TAPE & 0.122 & \textbf{\underline{109.023}} & 2.605 & \underline{1.56} \\
    & ProtT5 & \underline{0.190} & 106.776 & \underline{2.021} &  2.32 \\
    \midrule
    \multirow{4}{*}{$k$-modes} & 
     Seqvec & -0.093 & 0.910 & 1.078 & 18.29 \\
    & ESM2 & \underline{0.037} & \underline{47.333} & 5.461 &  2277.34 \\
    & TAPE & -0.116 & 1.035 & 0.984 &  \underline{8.96} \\
    & ProtT5 & -0.169 & 2.356 & \textbf{\underline{0.957}} &  9.50 \\
    \bottomrule
  \end{tabular}
  }
  \caption{Internal clustering quality metrics for $k$-means
    and $k$-modes on different embeddings on the
    Breast Cancer. Best values are shown in bold.}
  \label{tbl_clustering_Breast_Cancer}
\end{table}

\subsection{Overall Results - Observations}
Across all three datasets, we observe a pattern indicating the importance of geometry for downstream performance. The classification task aligns with a strong Poincaré hierarchy, in which ESM2 exhibits a better hyperbolic structure and consistently performs better on supervised tasks. Clustering aligns with Euclidean space, and ProtT5 exhibits the most separable Euclidean structure and gives better performance with Agglomerative and k-means clustering. Although it occasionally clusters well, SeqVec, the worst in classification, also has the most fragmented geometry, which explains its inconsistent behavior. Although TAPE exhibits a strong hyperbolic hierarchy, it does not translate into downstream performance, indicating that the hyperbolic structure alone is insufficient. In general, ESM2 excels at classification, ProtT5 at clustering, and SeqVec and TAPE exhibit weaker or less consistent geometric patterns.

\section{Conclusion}
\label{sec_conclusion}
In this paper, we present a geometric framework for evaluating LLM embeddings using $\delta$-hyperbolicity, ultrametricity, Neighbor Joining, and complementary measures such as Ricci curvature, persistent homology, and fractal dimension. We found that the embedding geometry aligns with the downstream performance. We found that ESM2's hyperbolic structure supports strong classification performance, that ProtT5’s Euclidean separability explains its superior clustering, and that weaker models like SeqVec show fragmented geometric patterns consistent with lower task performance. The results highlight that no single embedding is universally optimal. Each model excels in different geometric regimes. Each model is better suited to a specific task, and this structural analysis can pave the way towards guided, informed designs and the selection of LLMs.
Our analysis of synthetic spaces further confirms how hyperbolicity varies across geometric settings, emphasizing the importance of understanding embedding structure. All things considered, geometric metrics provide a useful perspective for choosing and fine-tuning LLM embeddings. Future work can extend this framework to broader datasets and explore integrating geometric insights into model design and training.

\bibliographystyle{plainnat-revised}
\bibliography{references}

\end{document}